\documentclass[11pt]{article}
\usepackage[cmex10]{amsmath}
\usepackage{amssymb}
\usepackage{graphicx}
\usepackage{mdwlist}
\usepackage{algorithmic}
\newtheorem{theorem}{Theorem}[section]

\newtheorem{definition}[theorem]{Definition}

\newcommand{\sq}{\hbox{\rlap{$\sqcap$}$\sqcup$}}
\newcommand{\qed}{\hspace*{\fill}\sq}

\newcommand{\prf}[1]{{}}

\newtheorem{Def}{Definition}[section]

\newcommand{\sacode}[5]
{ \vspace{.06in} \hrule \vspace{.06in} 
 \noindent {\bf #1}: \\
 \footnotesize \noindent {\bf Signature:}\B \nobreak
 \normalsize \begin{quote} \nobreak #2 \end{quote}
 \footnotesize \noindent {\bf States:}\B \nobreak
 \begin{quote} \nobreak #3 \end{quote}
 \noindent {\bf Transitions:} \nobreak
 \vspace{-.2in} \nobreak
 \normalsize #4
 \vspace{-.06in} \hrule \vspace{.06in} 
}

\newcommand{\act}[1]{%
    \relax\ifmmode
        \mathord{\mathcode`\-="702D\sf #1\mathcode`\-="2200}%
    \else
        $\mathord{\mathcode`\-="702D\sf #1\mathcode`\-="2200}$%
    \fi
}

\newcommand{\tup}[1]{%
    \relax\ifmmode
      \langle #1 \rangle%
    \else
        $\langle$#1$\rangle$%
    \fi
}

\newcommand{\seq}[1]{%
    \relax\ifmmode
      \langle \! \langle #1 \rangle \! \rangle%
    \else
        $\langle \! \langle$ #1 $\rangle \! \rangle$%
    \fi
}

\newcommand{\B}{\vspace*{-\smallskipamount}}

\newcommand{\ms}[1]{%
    \relax\ifmmode
        \mathord{\mathcode`\-="702D\it #1\mathcode`\-="2200}%
    \else
        {\it #1}%
    \fi
}

\newcommand{\lit}[1]{%
    \relax\ifmmode
        \mathord{\mathcode`\-="702D\sf #1\mathcode`\-="2200}%
    \else
        {\it #1}%
    \fi
}

\newcommand{\XDK}[1]{}
\newcommand{\remove}[1]{} 
\newcommand{\uselater}[1]{} 

\renewcommand{\setminus}{-}

\newcommand{\srvr}{s}

\newcommand{\qs}{\mathbb{Q}}
\newcommand{\quo}{Q}

\newcommand{\wSet}{\mathcal{W}}
\newcommand{\rdSet}{\mathcal{R}}

\newcommand{\srvSet}{\mathcal{S}}
\newcommand{\op}{{\pi}}
\newcommand{\rd}{{\rho}}
\newcommand{\wrt}{\omega}
\newcommand{\bef}{\rightarrow}

\newcommand{\tg}[1]{tag_{#1}}

\newcommand{\maxtag}{maxTag}

\newcommand{\qv}[1]{qView(#1)}
\newcommand{\sfw}{{\sc Sfw}} 
\newcommand{\aprxsfw}{{\sc Aprx-Sfw}} 
\newcommand{\cwfr}{{\sc CwFr}} 
\newcommand{\simple}{{\sc Simple}} 

\newcommand{\inter}[1]{I_{#1}}
\newcommand{\inps}[1]{\ms{inprogress}_{#1}}
\newcommand{\inp}[2]{\inps{#1}(#2)}

\newcommand{\mkSI}[1]{$#1${\sc -Set-Intersection}}

\newcommand{\setC}{{\sc Set-Cover}}

\begin{document}

\begin{titlepage}

\title{On the Practicality of Atomic MWMR Register Implementations.
\thanks{This work is supported by the Cyprus Research Promotion Foundation's grant
$\Pi$ENEK/0609/31 and the European Regional Development Fund.}}

\author{Nicolas C. Nicolaou
                                \thanks{Department of Computer Science,
                                        University of Cyprus, Cyprus. Email: \{nicolasn, chryssis\}@cs.ucy.ac.cy.}
                        \and Chryssis Georgiou$^\dag$      
      }

\date{}

\maketitle

\begin{abstract}
Multiple-writer/multiple-reader (MWMR) read/write atomic register implementations provide precise consistency guarantees, 
in the asynchronous, crash-prone, message passing environment. Fast MWMR atomic register 
implementations were first introduced in \cite{EGMNS09}. 
Fastness in their context was measured in terms of the number of single round read and write 
operations that does not sacrifice correctness. The authors in \cite{GNRS11} however, 
showed that decreasing the communication cost is not enough in these implementations.
In particular, considering that the performance is measured
in terms of the latency of read and write operations due to both (a) \emph{communication delays} and (b)
\emph{local computation}, they introduced two new algorithms that traded 
communication for reducing computation. As computation is still part of the algorithms, someone
may wonder: What is the trade-off between communication and local computation in real-time systems? 

In this work we conduct an experimental performance evaluation of four MWMR atomic register 
implementations: \sfw{} from \cite{EGMNS09}, \aprxsfw{} and \cwfr{}
from \cite{GNRS11}, and \simple{} (the generalization of \cite{ABD96} in the MWMR environment).  
We implement the algorithms on {\bf NS2}, a single processor simulator, and on {\bf PlanetLab},
a planetary-scale real-time network platform.
Due to its simplistic nature, \simple{} requires two communication round-trips per read or write operation,
but almost no local computation. The rest of the algorithms are (to this writing) the only to 
allow \emph{single} round read and write operations but require \emph{non-trivial} computation 
demands. We compare these algorithms with \simple{} and amongst each other to study the 
trade-offs between communication delay and local computation.
Our results shed new light on the
practicality of atomic MWMR register implementations.   

\vspace{2em}
\end{abstract}

\centerline{Technical Report TR-11-08}
\centerline{Department of Computer Science}
\centerline{University of Cyprus}
\centerline{September 2011}

\end{titlepage}

\section{Introduction}
\label{sec:intro}
\noindent{\bf\em Motivation and prior work:}
Frequent hardware failures increased the popularity of reliable distributed storage
systems as a medium for data availability and survivability; such systems 
are becoming increasingly important in data centers
and other enterprise settings~\cite{aguileratutorial}.
Traditional approaches achieve reliability by using a centralized controller 
to handle the replication of data over a redundant array of independent disks (RAID).
However, the centralized controller, which resides on a single location and 
is usually connected to the network via a single interface, constitutes 
a single point of failure and compromises data accessibility~\cite{CKGV08}. 

A distributed storage system overcomes this problem by replicating 
the data in geographically dispersed nodes, and ensuring 
data availability even in cases of complete site disasters. 
Distribution of data introduces, however, the problem of preserving 
data consistency between the replicas. Atomicity is the strongest consistency guarantee 
and provides the illusion that operations on the distributed storage
are invoked sequentially, even though they can be invoked concurrently.
The challenge of providing atomic consistency becomes even greater in the
presence of asynchrony and failures. In order to 
address this challenge and provably identify the trade-offs between consistency, reliability
and efficiency, researchers over the last two decades have focused on 
the simplest form of consistent distributed storage: \emph{an atomic read/write register}; 
for a recent survey see~\cite{attiyasurvey}. 
Atomic read/write register implementations can be
used directly to build distributed file systems (e.g.,~\cite{FL03}) and also
as building blocks for more complex distributed storage systems; for examples of
storage systems built on register implementations see~\cite{GL03, HPFAB04, Dynamo}. 

In this work we consider Multiple-Writer, Multiple-Reader (MWMR) atomic registers
over an asynchronous, crash-prone, message-passing setting.  
In such settings the register is replicated among a set of replica hosts
(or servers) to provide fault-tolerance and availability. Read and write
operations are implemented as communication protocols that ensure 
atomic consistency. 

The efficiency of register implementations is normally measured in terms 
of the latency of read and write operations. 
Two factors affect operation latency: 
(a) computation, and (b) communication delays. 
An operation communicates with servers to read
or write the register value. This involves at least 
a single communication round-trip, or \emph{round}, 
i.e., messages from the invoking process to some servers and 
then the replies from these servers to the invoking process.
Previous works focused on reducing the number of rounds
required by each operation.
Dutta et al.~\cite{CDGL04} developed the first 
single-writer/multi-reader (SWMR) algorithm, where all 
operations complete in a single round. 
Such operations are called \emph{fast}. 
The authors showed that fast operations are possible
 only if the number of readers in the system is constrained with respect 
 to the number of servers. They also showed that it is impossible
 to have MWMR register implementations where {\em all}
 operations are fast. 
 To remove the constraint on the number of readers, Georgiou et al.~\cite{GNS09} 
introduced  \emph{semifast} implementations where at most one complete two-round
read operation is allowed per write operation. They also showed that 
 semifast MWMR register implementations are impossible.

Algorithm \sfw, developed
  by Englert et al.~\cite{EGMNS09},  was the first to 
allow both reads and writes to be fast in the MWMR setting. 
The algorithm used {\em quorum systems}, sets of  intersecting subsets of servers, to handle server failures. 
  To decide whether an operation could terminate 
  after its first round, the algorithm employed two {\em predicates}, one for the write and one
  for read operations. 
   
   A later work by Georgiou et al. \cite{GNRS11} identified 
two weaknesses of algorithm \sfw{} with respect to its
practicality: (1) the computation required for calculating the predicates is NP-complete, and
 (2) fast operations are possible only when every \emph{five} or more quorums 
 have a 
 non-empty intersection. To tackle these issues the authors introduced two new algorithms. 
 The first algorithm, called \aprxsfw{}, proposed a polynomial $\log$-approximation 
 solution for the computation of the predicates in \sfw{}. This would allow faster computation 
 of the predicates while potentially increasing the number of two-round operations. 
To tackle the second weakness of \sfw{}, the authors 
presented algorithm \cwfr{} that uses  \emph{Quorum Views}  \cite{GNS08}, 
client-side decision tools, to allow some 
    fast {\em read} operations without additional constraints on the underlying quorum system. 
    Write operations in this implementation always take  two rounds to complete. 
    
All the above algorithms have been rigorously proven to guarantee atomic
consistency and theoretical analyses provide clues for their efficiency.
However, little work has been done to study the practicality and effectiveness of these approaches
over realistic, planetary-scale adverse environments. The efficiency of these approaches would
greatly affect any distributed file or storage system built on top of them. 
Furthermore, as the effect of the performance metrics is orthogonal one may 
wonder: "When shall we prefer reduced communication for increased computation and when vice-versa?"

    
\noindent{\bf\em Background.}
 Attiya et al.~\cite{ABD96} developed a SWMR algorithm that achieves consistency
by using intersecting majorities of servers in combination
with $\tup{\ms{timestamp},\ms{value}}$ value tags.
%
A write operation increments the writer's local timestamp
and delivers the new tag-value
pair to a majority of servers, taking one round.
A read operation obtains tag-value pairs
from some majority, then propagates the pair corresponding
to the highest timestamp to some majority of servers,
thus taking two rounds.

The majority-based approach in~\cite{ABD96}
is readily generalized to quorum-based approaches 
in the MWMR setting
(e.g.,~\cite{LS97,ES00,LS02,FL03,GAV07}).
Such algorithms requires at least two communication 
rounds for each read and write operation. Both 
write and read operations query the servers for the 
latest value of the replica during the first round. 
In the second round the write operation generates a new
tag and propagates the tag along with the new value 
to a quorum of servers. A read operation propagates to a quorum 
of servers the largest value it discovers during its first round.  
This algorithm is what we call \simple{} in the rest of this paper. 

Dolev \emph{et al.}~\cite{DGLSW03} 
and Chockler \emph{et al.}~\cite{CGGMS09},
provide MWMR implementations 
where some reads involve a single communication
round when it is confirmed that the value read was
already propagated to some quorum. Although, these
algorithm share some similarities with \cwfr{}, the latter
algorithm allows read operations to be fast even when 
those are concurrent with write operations.

Dutta et al.~\cite{CDGL04} present the first \emph{fast} atomic SWMR
implementation where all operations
take a {\em single} communication round.
They show that fast behavior is achievable only
when the number of reader processes $R$ 
is inferior to $\frac{S}{t}-2$,
where $S$ the number of servers, $t$ of whom may crash.
They also showed that fast MWMR implementations are
impossible even in the presence of a single server failure.
Georgiou et al.~\cite{GNS09}
introduced the notion of \emph{virtual nodes} that enables
an unbounded number of readers.
They define the notion of \emph{semifast} implementations
where only a single read operation per write needs to be ``slow''
(take two rounds). 
They also show the impossibility of  semifast MWMR implementations.

Georgiou et al.~\cite{GNS08} showed that fast and semifast
quorum-based SWMR implementations
are possible if and only if a common
intersection exists among all quorums. Hence
a single point of failure exists in such solutions
(i.e., any server in the common intersection),
making such implementations not fault-tolerant.
To trade efficiency for improved fault-tolerance,
{\em weak-semifast} implementations in~\cite{GNS08}
require at least one single slow read per write operation,
and where all writes are fast.
To obtain a weak-semifast implementation 
 they introduced a client-side
decision tool called \emph{Quorum Views} that enables fast
read operations under read/write concurrency 
when \emph{general quorum systems} are used.
    
 Recently, Englert \emph{et al.}~\cite{EGMNS09}  developed 
 an atomic MWMR register implementation, called algorithm \sfw{},
that allows both reads and writes to complete in a \emph{single  round}.
To handle server failures, their algorithm uses \emph{$n$-wise quorum systems}:  
a set of subsets of servers, such that each $n$ of 
these subsets intersect.  The parameter $n$ is called the \emph{intersection 
degree} of the quorum system.
The algorithm relies on $\tup{\tg{},value}$ pairs to totally order 
write operations.  In contrast with traditional approaches, the algorithm
uses the \emph{server side ordering} (SSO) approach that transfers
the responsibility of incrementing the tag from the writers to the servers.
This way, the \emph{query} round of write operations is eliminated. 
The authors proved that fast MWMR implementations are possible if and only if they allow
not more than $n-1$ successive write operations, where $n$ is the intersection 
degree of the quorum system. If read operations are also allowed to modify 
the value of the register then from the provided bound it follows that a 
fast implementation can accommodate up to $n-1$ readers and writers. 

\noindent{\bf\em Contributions.}
In this work we attempt to provide empirical evidence on the efficiency 
and practicality of MWMR atomic register implementations and give an answer to the above question. More
precisely, we experimentally evaluate the performance of algorithms \sfw{}, \aprxsfw{},
and \cwfr{} as they are the only known algorithms to this writing to allow 
single round writes and reads in the MWMR model. To observe the benefits of reducing the number 
of communication rounds per operation, we compare our findings with the performance of 
algorithm \simple{}, an all two-round algorithm. To test the \emph{efficiency} of the 
algorithms we first implement them on NS2 \cite{NS2}, and to test their {\em practicality} 
we deploy them on PlanetLab \cite{planetlab}, a planetary scale real-time network platform. 
In detail:

\begin{enumerate}

\item To test \emph{efficiency},  we simulate the algorithms 
on the NS2 \cite{NS2} network simulator. The controlled environment of NS2 allows us to test the 
performance of the algorithms under different environmental conditions.  
In particular, the operation latency of the algorithms is tested under the following simulation scenarios:
(1) Variable number of readers/writers/servers, (2) Deployment of different quorum constructions,
and (3) Variable network delay. 

Our preliminary results on NS2 confirm the high computation demands of \sfw{} over
\aprxsfw{} as theoretically proven by \cite{GNRS11}. In addition, they suggest that the 
			computation burden needed by \aprxsfw{} and \cwfr{} was lower than the communication 
			cost of a second communication round in most scenarios, by comparing the two algorithms to algorithm \simple{}. 
			In terms of scalability, it appears that every algorithm suffers a performance degradation 
			as the number of participants is increasing. 
			Finally, we observe that long network delays promote algorithms with high computational demands and 
			fewer communication rounds (such as algorithm \aprxsfw). 
			In general we can say that the simulation promotes \cwfr{} as the most efficient algorithm,
			in most of the scenarios.

\item To test \emph{practicality}, we implement and deploy our algorithms on PlanetLab \cite{planetlab}, an overlay network infrastructure 
composed of machines that are located throughout the globe.
Given the adverse and unpredictable conditions of the real-time system,
we measure and compare the operation latency of the algorithms under two different  
families of service scenarios:
(1) Variable number of readers/writers/servers, (2) Deployment of different quorum constructions.
In our implementations communication was established via TCP/IP and the C/C++ programming language 
and sockets were used for interfacing with TCP/IP.  

Our findings also suggest that the computation burden of algorithms \cwfr{} and \aprxsfw{} 
is lower than the communication cost of the second round required by algorithm \simple{} 
in most of the scenarios. 
More precisely, computation does not appear to have a great impact on the performance
of the algorithms. 
This is partly due to the fact that both \cwfr{} and \aprxsfw{}
exhibit a percentage of slow operations under 20\%. Also, unlike NS2, there are 
a number of machines executing our protocols and thus computation is no longer 
performed by a single processor. In terms of scalability, we still observe a degradation
on the performance of the algorithms as the number of participants increases. 
In addition, we observe that the intersection degree
of the quorum system can play a decisive factor as it affects in a large degree 
the performance of \aprxsfw{}. The higher the intersection degree the more reads/writes
can be fast. Even though this is true, the average latency 
achieved by \aprxsfw{} in environments with small intersection degree is not 
much higher than the latency of the competition.  
In general we can say that PlanetLab promotes \aprxsfw{} as the most practical algorithm,
in most of the scenarios. 
\end{enumerate}

Combining our findings in both experimentation environments, we may 
generally conclude that in most runs, algorithms \aprxsfw{} and \cwfr{} perform better than 
algorithm \simple{}. This suggests that the additional computation 
incurred in these two algorithms does not exceed the delay
associated with a second communication round. Furthermore, depending on the server
configuration used and the frequency of read and write operations, we draw a clear
line when algorithm \cwfr{} should be preferred over algorithm \aprxsfw{} and vice-versa.

\paragraph{Paper organization.}
In Section \ref{sec:model} we briefly describe the model of computation 
that is assumed by the implemented algorithms. 
In Section \ref{sec:alg} we provide a high level description of the 
algorithms we examine.
In Section~\ref{sec:ns2} we overview NS2, we present our testbed and
we provide the scenarios we consider. The results from our NS2 simulation 
are presented in Section~\ref{sec:ns2res}. 
In Section~\ref{sec:pl} we overview PlanetLab, and we present our testbed,
the scenarios we consider and briefly mention the implementation
difficulties we encountered.  
PlanetLab results are given in Section~\ref{sec:plres}. 
We conclude in Section \ref{sec:conc}.

\section{Model and Definitions}
\label{sec:model}

We consider the asynchronous 
message-passing model.
There are three distinct finite sets of crash-prone
processors:
a set of readers $\rdSet$, a set of writers $\wSet$, and a set of 
servers $\srvSet$ .
The identifiers of all processors are unique and comparable. 
Communication among the processors is accomplished 
via reliable communication channels.  

\paragraph{Servers and quorums.}
Servers are arranged into intersecting sets, or    
\emph{quorums}, that together form a quorum system~$\qs$. 
For a set of quorums $\mathcal{A}\subseteq \qs$ we denote
the intersection of the quorums in $\mathcal{A}$ by 
$\inter{\mathcal{A}}=\bigcap_{Q\in \mathcal{A}}Q$.
A quorum system $\qs$ is called an \emph{n-wise quorum system} if 
for any $\mathcal{A}\subseteq \qs$, s.t. $|\mathcal{A}|=n$ we have
$\inter{\mathcal{A}}\neq \emptyset$.
We call $n$ the \emph{intersection degree} of $\qs$.
Any quorum system is  
a {\em 2-wise} (pairwise) quorum system
because any two quorums intersect.
At the other extreme, a \emph{$|\qs|$-wise} quorum system has 
a common intersection among all quorums.  
From the definition it follows that an \emph{n-wise} quorum system 
is also a \emph{k-wise} quorum system, for $2\leq k \leq n$.  

Processes may fail by crashing.
A process $i$ is \emph{faulty} in 
an execution 
if $i$ crashes in 
the execution (once a process crashes, it does not recover);
otherwise $i$ is \emph{correct}.
A quorum $Q\in\qs$ is non-faulty if $\forall i\in Q$, $i$ is correct;
otherwise $Q$ is faulty. 
We assume that
at least one quorum in $\qs$ is non-faulty in any execution.

\paragraph{Atomicity.}
We study atomic read/write register implementations, 
where the register is replicated at servers.
Reader $p$ requests a read operation $\rd$ on the register 
 using action $\act{read}_{p}$. Similarly, 
a write operation is requested using action
$\act{write}(*)_{p}$ at writer $p$.
The steps corresponding to such actions are called
\emph{invocation} steps.
An operation terminates with the corresponding 
acknowledgment
action;
these steps 
are called
\emph{response} steps.
An operation $\op$ is \emph{incomplete} in an execution 
when the invocation step of $\op$ does not have the 
associated response step; otherwise $\op$ is 
\emph{complete}. 
We assume that requests made by read and write processes 
are {\em well-formed}: a process
does not request a
new operation until it receives the response for
a previously invoked operation.

In an execution, we say that an operation (read or write) $\op_1$ {\em 
precedes} another operation $\op_2$, or $\op_2$ {\em succeeds} $\op_1$, 
if the response step for $\op_1$ precedes in real time the invocation 
step of $\op_2$; this is denoted by $\op_1\bef\op_2$.  Two operations 
are {\em concurrent} if neither precedes the other.

Correctness of an implementation of an atomic read/write object 
is defined in terms 
of the {\em atomicity} and {\em termination}
properties. Assuming the
failure model discussed earlier, the termination property requires 
that any operation invoked by a correct process eventually completes.  
Atomicity is defined as follows~\cite{Lynch1996}.
For any execution 
if all read and 
write operations that are invoked complete, then the
operations can be partially ordered by an ordering $\prec$, so that the 
following properties are satisfied:
	\begin{itemize*}
		\item [\em P1.] The partial order is consistent with the 
					external order of invocation and responses, that is, there do 
					not exist operations $\pi_1$ and $\pi_2$, 
					such that $\pi_1$ completes before $\pi_2$ starts, 
					yet $\pi_2 \prec \pi_1$.
		\item[\em P2.] All write operations are totally 
					ordered and every read operation is ordered with respect 
					to all the writes.
		\item[\em P3.] Every read operation ordered after any writes returns
the value of the last write preceding it in the partial order, and any
read operation ordered before all writes returns the initial value
of the register.
	\end{itemize*}
%
%
	
\paragraph{Efficiency and Fastness.}
We measure the efficiency of an atomic 
register implementation in terms of \emph{computation} and 
\emph{communication round-trips} (or simply rounds). 
 A round is defined as follows~\cite{CDGL04,GNS09,GNS08}:
\begin{definition}\label{def:com}
Process $p$ performs a communication round during operation $\op$ 
if all of the following hold:
1. $p$ sends request messages that are a part of $\op$ 
         to a set of processes,
2. any process $q$ that receives a request message from $p$ for 
         operation $\op$, replies
without delay.
3. when process $p$ receives enough replies it 
terminates the round (either completing $\op$
or starting new round).
\end{definition}

Operation $\op$ is \emph{fast} \cite{CDGL04} if it completes after 
its first communication round;
an implementation is fast if in each execution
all operations are fast.
We use quorum systems 
and tags to maintain, and impose an ordering on, the 
values written to the register replicas.
We say that a 
quorum $Q\in\qs$, \emph{replies} 
to a process $p$ for an operation $\op$ during a round,
if $\forall s\in Q$, $s$ receives a message during the round
and replies to this message, and $p$ receives all such replies.

\remove{
Given that any subset of readers or writers may crash, the 
termination of an operation cannot depend on the progress
of any other operation.
Furthermore 
we guarantee termination only if servers' replies within
a round of some operation do not depend on receipt 
of any message sent by other processes.
%
Thus we can construct executions where only the messages from the
invoking processes to the servers, and from the servers to the invoking 
processes are delivered.
Lastly, to guarantee termination under the assumed failure model, 
no operation can wait for more than a singe quorum to reply
within the processing of a single round.
}

\section{Algorithm Description}
\label{sec:alg}

Before proceeding to the description of our experiments we first present a high level description of the
algorithms we evaluate.
We assume that the algorithms use quorum systems and follow the failure
model presented in Section \ref{sec:model}. Thus, termination is guaranteed if any read and write operation waits from
the servers of a single quorum to reply. To order the written values the algorithms use (tag, value) pairs,
where a tag contains a timestamp and the writer's identifier.

\subsection{Algorithm \simple{}}
Algorithm Simple is a generalization of the algorithm developed by Attiya et al. \cite{ABD96} for the MWMR
environment. Servers run the Server protocol, writers the Write protocol, and readers the Read protocol,
as described below:\vspace{.3em}

\noindent{\bf Server Protocol:} Each replica receives read and write requests, and updates its local copy of the
replica if the tag enclosed in the received message is greater than its local tag before replying with an
acknowledgment and its local copy to the requester.\vspace{.3em}

\noindent{\bf Write Protocol:} The write operation performs two communication rounds. In the first round the
writer sends query messages to all the servers and waits for a quorum of servers to reply. During the
second round the writer performs the following three steps: (i) it discovers the pair with the maximum tag
among the replies received in the first round, (ii) it generates a new tag by incrementing the timestamp
inside the maximum discovered tag, and (iii) it propagates the new tag along with the value to be written
to a quorum of servers.\vspace{.3em}

\noindent{\bf Read Protocol:} Similarly to the write operation every read operation performs two rounds to
complete. The first round is identical as the first round of a write operation. During the second round
the read operation performs the following two steps: (i) it discovers the pair with the maximum tag
among the replies received in the first round, and (ii) it propagates the maximum tag-value pair to a
quorum of servers.

\subsection{Algorithm \sfw{}}
Algorithm \sfw{} assumes that the servers are arranged in an $n$-{\em wise} 
quorum system. To enable fast writes the algorithm  assigns partial 
responsibility to the servers for the ordering of the values written.
Due to concurrency and asynchrony, however,
two servers may receive messages originating from two different writers in different 
order. Thus, a read or write operation may witness different tags assigned to a single write 
operation. To deal with this problem, 
algorithm \sfw{} uses two {\em predicates} to determine
whether ``enough'' servers in the replying quorum 
assigned the same tag to a particular write operation.\vspace{.3em} 

\noindent{\bf Server Protocol:}
Servers wait for read and write requests. When a server receives a write request
it generates a new tag, larger than any of the tags it witnessed, and assigns it to the value enclosed in
the write message. The server records the generated tag, along with the write operation it was created
for, in a set called $inprogress$. The set holds only a single tag (the latest generated by the server) for
each writer.\vspace{.3em}

\noindent{\bf Write Protocol:}
Each writer must communicate with a quorum
of servers, say $Q$, during the first round of each 
write operation.At the end of the first round the writer evaluates a predicate to determine
whether "enough" servers replied with the same tag. 
Let $n$ be the intersection degree
of the quorum system, and $\inp{s}{\wrt}$ be the $inprogress$ set 
that server $s$ enclosed in the message it sent to the writer that invoked $\wrt$.
The write predicate is:
\vspace{.2cm}

\noindent{\bf PW: Writer predicate for a write $\wrt$:} 
		$\exists\ \tau, A, MS$ where: 
		$\tau\in \{\tup{.,\wrt}:\tup{.,\wrt}\in \inp{s}{\wrt}~\wedge~s\in Q\}$,
    $A\subseteq\qs,0\leq|A|\leq\frac{n}{2}-1$, and
    $MS=\{s:s\in Q~\wedge~\tau\in\inp{s}{\wrt}\}$, 
    s.t. either $|A|\neq 0$ and $\inter{A}\cap Q\subseteq MS$ or $|A|=0$ and $Q=MS$.
\vspace{.2cm}

The predicate examines whether the same tag for the ongoing 
write operation is contained in the replies of all servers 
in the intersection among the replying quorum and $\frac{n}{2}-1$ other quorums.
Satisfaction of the predicate for a tag $\tau$
guarantees that any subsequent operation will also determine that the write operation is 
assigned tag $\tau$. 
If the predicate {\bf PW} holds 
then the write operation is fast. Otherwise the writer 
assigns the highest tag to the written value and proceeds to a second round to propagate the highest discovered tag to a quorum of servers. \vspace{.3em}

\noindent{\bf Read Protocol:} Read operations take one or two rounds. During its first round the read collects replies
from a quorum of servers. Each of those servers reports
a set of tags (one for each writer). The reader 
needs to decide which of those tags is assigned to 
the latest potentially completed write operation.
For this purpose it uses a predicate similar to {\bf PW}:
 
\noindent{\bf PR: Reader predicate for a read $\rd$:} 
		$\exists\ \tau, B, MS$, where:
    $\max(\tau)\in \bigcup_{s\in Q}\inp{s}{\rd}$, 
    $B\subseteq\qs,0\leq|B|\leq\frac{n}{2}-2$, and
    $MS=\{s:s\in Q~\wedge~\tau\in\inp{s}{\rd}\}$,
    s.t. either $|B|\neq 0$ and $\inter{B}\cap Q\subseteq MS$ or $|B|=0$ and $Q=MS$.
\vspace{.2cm}

The predicate examines whether there is a tag for 
some write operation that is contained in the replies of all servers 
in the intersection among the replying quorum $\frac{n}{2}-2$ other quorums.
Satisfaction of the predicates for a tag $\tau$
assigned to some write operation, guarantees that any subsequent operation will also determine that the write operation is assigned tag $\tau$. 
A read operations can be fast even 
if {\bf PR} does not hold, but the read observed enough $confirmed$ tags with the same value. 
Confirmed tags are maintained in the servers and they indicate 
that either the write of the value with that tag is complete, or the tag was returned by some read operation.\vspace{.3em}

The interested reader can see~\cite{EGMNS09} for full details.

\subsection{Algorithm \aprxsfw{}}
The complexity of the predicates in \sfw{} raised the question whether
they can be computed efficiently. In a recent work~\cite{GNRS11} (see also~\cite{D3-TR})
we have shown that both predicates are NP-Complete. To prove the NP-completeness of the predicates,
we introduced a new combinatorial problem, called \mkSI{k}, which 
captured both {\bf PW} and {\bf PR}. An approximate solution to the new problem
could be obtained polynomially by using the approximation algorithm for the 
set cover. 
The steps of the approximation algorithm are:

\begin{figure}[!ht]
\hrule\vspace{.15cm}
{\small
{\raggedright
\noindent Given $(U,M,\qs,k)$: \\
\emph{Step 1:} $\forall m\in M$\\
~ ~ ~ ~ ~ let $T_m=\{(U\setminus M) \setminus (Q_i\setminus M): m\in Q_i\}$ \\
\emph{Step 2:} Run \setC\ greedy algorithm on \\
~ ~ ~ ~ ~ the instance $\{U\setminus M, T_m, k\}$ for every $m\in M$:\\ 
\hspace*{1em}\emph{Step 2a:} Pick the set $R_i\in T_m$ with \\
~ ~ ~ ~ ~ ~ ~ the maximum uncovered elements \\
\hspace*{1em}\emph{Step 2b:} Take the union of every $R\in T_m$ \\
~ ~ ~ ~ ~ ~ ~ picked in Step 2a (incl. $R_i$) \\
\hspace*{1em}\emph{Step 2c:} If the union equals $U\setminus M$ go to Step 3;\\
~ ~ ~ ~ ~ ~ ~ else if there are more sets in $T_m$ go to Step 2a\\
~ ~ ~ ~ ~ ~ ~ else repeat for another $m\in M$ \\ 
\emph{Step 3:}  For any set $(U\setminus M)\setminus (Q_i\setminus M)$ in the solution 
of set cover, add $Q_i$ in the intersecting set. \vspace{3pt}
}
}
\hrule
\caption{Polynomial approximation algorithm for the \mkSI{k}.}
\label{fig:apprx}
\end{figure}

By setting $U=\srvSet$, $M$ to contain 
all the servers that replied with a particular tag in the first 
round of a read or write operation, and $k$ to be
$\frac{n}{2}-1$ for {\bf PW} and $\frac{n}{2}-2$ for {\bf PR}, we obtain an approximate solution for \sfw{}. The new
algorithm, called \aprxsfw{}, inherits the read, write, and serve protocols of \sfw{} and uses 
the above approximation algorithm for the evaluation of the {\bf PW} and {\bf PR} 
predicates. \aprxsfw{} promises to validate the predicates only when \sfw{} validates the predicates
(preserving correctness), 
and yields a factor of $\log |\srvSet|$ increase on 
the number of second communication rounds.
This is a modest price to pay in exchange for substantial reduction
in the computation overhead of algorithm \sfw{}.

\subsection{Algorithm \cwfr{}}
A second limitation of \sfw{} is its reliance to specific
constructions of quorums to enable fast read and write operations.  
Algorithm \cwfr{}, presented in \cite{GNRS11} (see also~\cite{D2-TR}), is designed to overcome this limitation,
yet trying to allow single round read and write operations.
While failing to enable single round writes, \cwfr{} 
enables fast read operations by adopting the general idea of 
Quorum Views \cite{GNS08}. 
The algorithm employs two techniques:(i) the typical query and propagate approach (two rounds) for write operations, and 
(ii) analysis of Quorum Views \cite{GNS08} for potentially fast (single round) read operations.

Quorum Views are client side tools that, based on the distribution 
of a tag in a quorum, may determine the state of a write operation: 
completed or not. In particular, there are three different classes 
of quorum views. $\qv{1}$ requires that 
all servers in some quorum reply with the same tag revealing that
the write operation propagating this tag has potentially completed. 
$\qv{3}$ requires that some servers in the quorum contain an older value,
but there exists an intersection where all of its servers 
contain the new value. This creates uncertainty whether the write operation 
has completed in a neighboring quorum or not. Finally 
$\qv{2}$ is the negation of the other two views and requires a quorum where the new value
is neither distributed to the full quorum nor distributed fully
in any of its intersections. This reveals that the write operation 
has certainly not completed.

%
%
%

Algorithm \cwfr{} incorporates an iterative technique
around quorum views that not only predicts
the completion status of a write operation, but also detects the last potentially
complete write operation. Below we provide a description 
of our algorithm 
and present the main idea behind our technique.\vspace{.3em}

\noindent{\bf Write Protocol:}
The write protocol has two rounds. 
During the first round the writer 
discovers the maximum tag among the servers: 
it sends
read messages to all servers and waits for replies from all  members of 
some quorum. 
It then discovers the maximum tag among the replies and 
generates a new tag in which it encloses the incremented
timestamp of the maximum tag, 
and the writer's identifier.
In the second round, the writer associates the value to be written 
with the new tag, it propagates the pair 
to some quorum, and completes the write.\vspace{.3em}

\noindent{\bf Read Protocol:}
The read protocol is more involved. 
The reader sends
a read message to all servers and waits for some quorum to reply.
Once a quorum replies, the reader
determines $\maxtag{}$. 
%
Then the reader analyzes the distribution of the tag within the 
responding quorum $\quo{}$ in an attempt to determine 
the latest, potentially complete, write operation.
%
This is accomplished by determining the quorum view conditions.
Detecting conditions of $\qv{1}$ and
$\qv{3}$ are straightforward. When condition for $\qv{1}$ is 
detected, the read completes and the value associated with the discovered
$\maxtag{}$ is returned.
In the case of $\qv{3}$ the reader continues to the
second round, advertising the latest tag ($\maxtag{}$) and its associated value.
When a full quorum replies in the second round, the 
read returns the value associated with $\maxtag{}$.
Analysis of $\qv{2}$ involves the discovery of the earliest completed write
operation. This is done iteratively by (locally) removing the servers 
from $Q$ that replied with the largest tags.  
After each iteration 
the reader determines the next largest tag 
in the remaining server set,
and then re-examines the quorum views in the next iteration. 
This process eventually leads to either $\qv{1}$ or $\qv{3}$
being observed.
If $\qv{1}$ is observed, 
then the read completes in a single round by returning the 
value associated with the maximum tag 
among the servers that \emph{remain} 
in $Q$. 
If $\qv{3}$ is observed, 
then 
the reader proceeds to the second round as above,
and upon completion it returns the value associated with
the maximum tag $\maxtag{}$ discovered among the original respondents in $\quo{}$.\vspace{.3em}

\noindent{\bf Server Protocol:}
The servers play a passive role.
They 
receive 
read or write requests, 
update their object replica
accordingly,
and 
reply to the process that invoked the operation. 
Upon receipt of any message, the 
server compares its local tag with the tag included in the message.
If the tag of the message is higher than its local tag, the 
server adopts the higher tag along with its corresponding 
value. Once this is done the server replies
to the invoking process.

\subsection{Algorithm Overview}
Table \ref{tab:rev} accumulates the theoretical communication
and computation burdens of the four algorithms we consider.
The name of the algorithm appears in the first column of the table. The second and third columns
of the table shows how many rounds are required per write and read operation respectively. 
The next two columns present the computation required by each algorithm and the last column
the technique the algorithm incorporates to decide on the values read/written on the atomic
register. 

\begin{table}[h]
	\begin{center}
		\begin{tabular}{|l|c|c|c|c|c|}
			\hline7
			{\bf Algorithm} & {\bf WR} & {\bf RR} & {\bf RC} & {\bf WC} & {\bf Decision Tool} \\
			\hline 
			\hline 
			\simple{} & 2 & 2 & $O(|\srvSet|)$ & $O(|\srvSet|)$ & Highest Tag \\
			\hline
			\sfw{} & 1 or 2 & 1 or 2 & $O(2^{|\qs|-1})$ & $O(2^{|\qs|-1})$ & Predicates \\
			\hline
			\aprxsfw{} & 1 or 2 & 1 or 2 & $O(|\wSet||\srvSet|^2|\qs|)$ & $O(|\srvSet|^2|\qs|)$ & Predicate Approximation \\
			\hline
			\cwfr{} & 2 & 1 or 2 & $O(|\srvSet||\qs|)$ & $O(|\srvSet|)$ & Quorum Views / Highest Tag \\
			\hline
		\end{tabular}
	\end{center}\vspace{-1em}
	\caption{Theoretical comparison of the four algorithms.}
	\label{tab:rev}
\end{table}

\section{NS2-Simulation}
\label{sec:ns2}
In this section we describe in detail our NS2 simulations.
We provide some basic details about the NS2 simulator 
and then we present our testbed. Following 
our testbed, we present the parameters we considered and 
the scenarios we run for our simulations. 

\subsection{The NS2 Network Simulator}
\label{ssec:NS2}
NS2 is a discrete event network simulator \cite{NS2}. It is an open-source project built in C++ that allows users to extend 
its core code by adding new protocols. 
Because of its extensibility and plentiful online documentation, 
NS2 is very popular in academic research. Customization of the NS2 simulator 
allows the researcher to obtain full control over the event scheduler and the deployment environment. 
Complete control over the simulation environment and its components will help us investigate 
the exact parameters that are affected by the implementation of the developed algorithms. 
Performance of the algorithms is measured in terms of the ratio of the number of fast over slow R/W operations
(communication burden), 
and the total time it takes for an operation to complete (communication + computation = operation latency). 
Measurements of the performance 
involves multiple execution scenarios; each scenario is dedicated in investigating the behavior 
of the system affected by a particular system characteristic. The following system components will be 
used to generate a variety of simulation executions and enable a more comprehensive evaluation of the 
developed algorithms. Notice that each component affects a different aspect of the modeled environment. 
Thus, studying executions affected by the variation of a single or multiple components are both of great importance.  

\subsection{Experimentation Platform}
\label{ssec:ns2testbed}

Our test environment consists
of a set of writers, readers, and servers. 
Communication between the nodes is established via 
bidirectional links, with:
\begin{itemize}
	\item 1Mb bandwidth, 
	\item latency of $10ms$, and 
	\item DropTail queue.
\end{itemize}
To model local asynchrony, the processes send messages 
after a random delay between 0 and 0.3 $sec$. 
%
We ran NS2 in Ubuntu,
on a Centrino 1.8GHz processor. The average of 5 samples
per scenario provided the stated operation latencies.
 
We have evaluated the algorithms with majority quorums. 
 As discussed in \cite{EGMNS09}, assuming $|\srvSet|$ servers
 out of which $f$ can crash, we can construct an
 $(\frac{|\srvSet|}{f}-1)$-wise quorum system $\qs$. 
 Each quorum $Q$ of $\qs$ has size $|Q|=|\srvSet|-f$. 
 The processes are not aware of $f$.  The quorum system is 
generated \emph{a priori} and is distributed to each participant 
node via an external service (out of the 
scope of this work). 

We model server failures by selecting some quorum of servers 
(unknown to the participants) to be correct and allowing any other server to crash. 
The positive time parameter $cInt$ is used to model
the failure frequency or reliability of every server $\srvr$.
For our simulations we initialize $cInt$ to be equal to 
the one third of the total simulation time. Each time a server
checks for failure, it cuts $cInt$ in half until it becomes 
less than one. A failure is generated as following. First,
the server determines whether it belongs in the correct quorum.
If not the server sets its crash timeout to $cInt$. Once $cInt$ 
time is passed, the server picks a random number between 0 and 100.
If the number is higher than 95 then the server stops. In other 
words a servers has 5\% chance to crash every time the timer expires.

We use the positive time parameters $rInt=4sec$ and $wInt=4sec$
 to model operation frequency. Readers and writers pick a uniformly at random 
 time between $[0\ldots rInt]$ and $[0\ldots wInt]$, respectively,
 to invoke their next read (resp. write) operation. 
%

Finally we specify the number of operations each participant should 
invoke. For our experiments we allow participants to perform up to 25 
operations (this totals to 500-4000 operations in the system).

\subsection{Scenarios}
\label{ssec:ns2scenarios}

The scenarios were designed to test
(i) scalability of the algorithms as the number of 
readers, writers and servers increases, (ii) the relation between 
 quorum system deployment and operation latency, and (iii) whether 
network delays may favor the algorithms that minimize the communication rounds. 
In particular we consider the following parameters:
\begin{enumerate}

	\item {\bf Number of Participants:} We run every test with 10, 20, 40, and 80 readers and writers.
	To test the scalability of the algorithms with respect to the number
	of replicas in the system we run all of the above tests with 10, 15, 20, and 25 servers. 
	Such tests highlight whether an algorithm is affected by the number of participants in the system. 
	Changing the number of readers and writers help us investigate how each algorithm 
	handles an increasing number of concurrent read and write operations. The more the servers
	on the other hand, the more concurrent values may coexist in the service. So, algorithms 
	like \aprxsfw{} and \cwfr{}, who examine all the discovered values and do not rely on the 
	maximum value, may suffer from local computation delays.

	\item {\bf Quorum System Construction:} As mentioned in Section \ref{ssec:ns2testbed} we use
	majority quorums as they can provide quorum systems with high intersection degree.
	So, assuming that $f$ servers may crash we construct quorums of size 
	$|\srvSet|-f$. 
	As the number of servers $|\srvSet|$ varies between 10,15,20, and 25, we run the tests for 
	two different failure values, i.e. $f=1$ and $f=2$. This affects our environment in two ways:
	\begin{itemize}
		\item[(i)] We get quorum systems with different quorum intersection degrees. 
		According to \cite{EGMNS09} for every given $\srvSet$ and $f$ we obtain a $(\frac{|\srvSet|}{f}-1)$-wise quorum system.
		\item[(ii)] We obtain quorum systems with different number of quorum members. For example assuming
		15 servers and 1 failure we construct 15 quorums, whereas assuming 15 servers and 2 failures 
		we construct 105 different quorums. 
	\end{itemize}
	Changes on the quorum constructions help us evaluate how the algorithms
	handle various intersection degrees and quorum systems of various memberships. 
	
	\item {\bf Network Latency:} Operation latency is affected by local computation and communication 
	delays. As the speed of the nodes is the same, it is interesting to examine what is the impact
	of the network latency on the overall performance of the algorithms. In this scenario we examine
	whether higher network latencies may favor algorithms (like \cwfr{} and \aprxsfw{}) that, although 
	they have high computation demands, they allow single round operations. For this scenario 
	we change the latency of the network from 10ms to 500ms and we deploy a 6-wise quorum construction. 

\end{enumerate}

Another parameter we experimented with was operation frequency. Due to the invocation of operations 
in random times between the read and write intervals as explained in Section \ref{ssec:ns2testbed},
operation frequency varies between each and every participant. Thus, fixing different initial 
operation frequencies does not have an impact of the overall performance of the algorithms. For this
reason we avoided running our experiments over different operation frequencies.

\section{NS2 Results}
\label{sec:ns2res}
 
In this section we discuss our findings. 
First we compare the operation latency in 
algorithms \sfw{} with \aprxsfw{} to examine
our theoretical claims about the computational hardness of the
two algorithms. Then we compare algorithms \cwfr{}, \aprxsfw{}, and 
\simple{} to establish conclusions on the overall performance 
(including computation and communication) of the algorithms. 
We present a sample of plots that accompany our results. 
All the plots obtained by this experiment appear in \cite{D5-TR}. 

\subsection{Algorithm \sfw{} vs. \aprxsfw{}} 
\label{ssec:sfwvsaprx}

In~\cite{GNRS11} the authors showed that the predicates
used by \sfw{} were NP-complete. That motivated the introduction of
the \aprxsfw{} approximation algorithm, that could provide a polynomial, 
$\log$ approximation solution to the computation of the read and write 
predicates used in \sfw{}. To provide an experimental proof of the 
results presented in \cite{GNRS11} we implemented both algorithms 
in NS2. We run the two algorithms using different environmental 
parameters and we observed the latency of read and 
write operations in any of those cases. In particular we run scenarios with 
$|\rdSet|\in[10, 20, 40, 80]$ readers, with $|\wSet|=[10, 20, 40]$ writers, and with 
$|\srvSet|= [10, 15, 20]$ servers where $f=1$ may crash.
The exceedingly large delay of \sfw{} in scenarios with many servers and 
writers prevented us from obtaining results for bigger
$\srvSet$ and $\wSet$. The results we managed to obtain however were enough 
to reveal the difference between the two approaches.   

\begin{figure*}[!ht]
{\bf 20 Readers, 20 Writers, f=1:}\\	
	\begin{center}
	\vspace{-.7cm}
	\includegraphics[totalheight=2.3in]{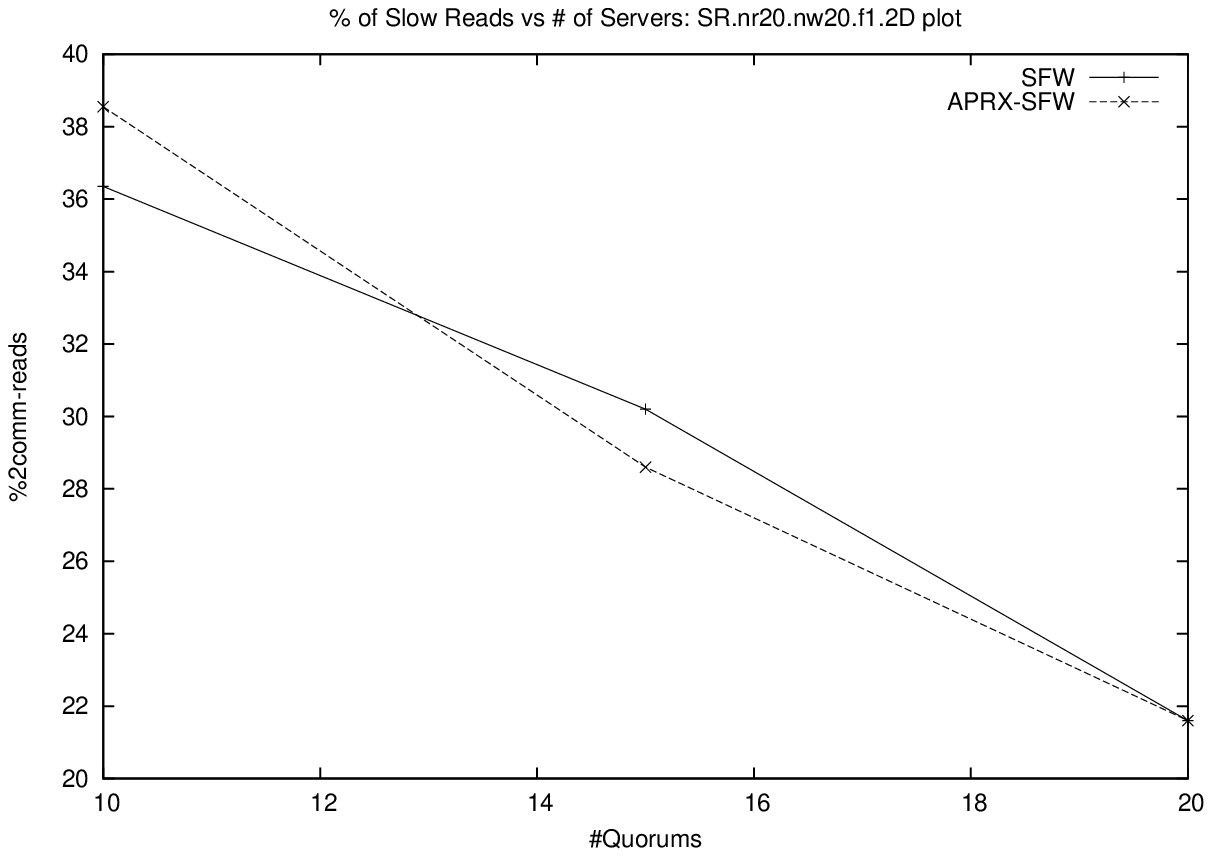}
	\hfill
	\includegraphics[totalheight=2.3in]{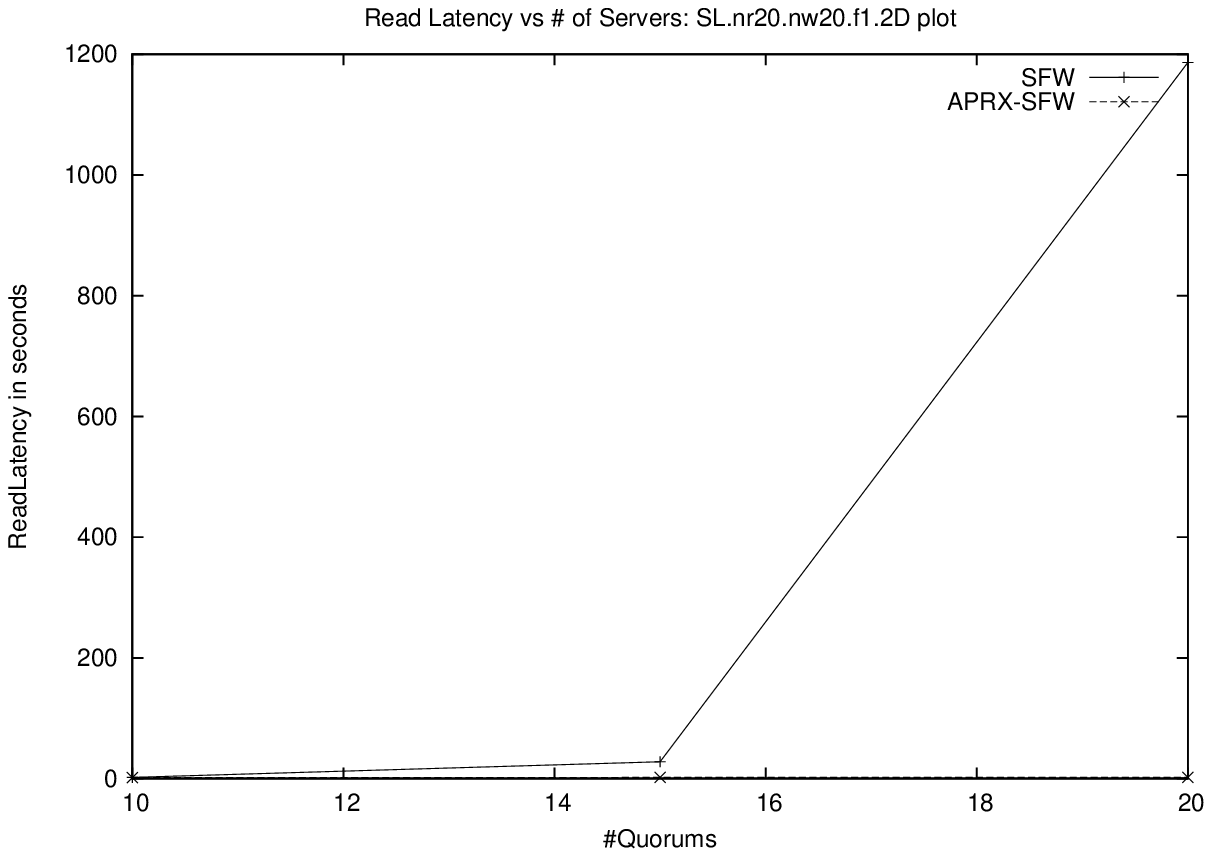}\\
	\end{center}
\caption{{\bf Left Column:} Percentage of slow reads,  {\bf Right Column:} Latency of read operations}	
	\label{fig:sfw}
\end{figure*}

Figure~\ref{fig:sfw} presents a specific scenario where $|\rdSet|=|\wSet|=20$.
Examining the latency of the two 
algorithms, including both communication and computation costs, provides
evidence of the 
heavy computational burden of algorithm \sfw{}. 
It appears that the average latency of the read operations (Figure~\ref{fig:sfw} right column)
in algorithm \sfw{} grows exponentially with respect to the number of servers (and thus quorum members) 
in the deployed quorum system. As it appears in the figure, the average latency 
for every read in \sfw{} was little lower than 200sec when using 15 quorums,
and then it exploded close to 1200sec when the number of quorums is 20. 
On the other hand the average latency of read operations in \aprxsfw{} 
grows very slowly. The average number of slow reads as it appears on the 
left plot of Figure~\ref{fig:sfw} shrinks as the number of quorums grows,
for both algorithms. Notice that as the number of servers grows the 
intersection degree which is equal to $n=(\frac{|\srvSet|}{f}-1)$ grows as well
since $f$ is fixed to $1$. From the figure we also observe that although 
the approximation algorithm may invalidate the predicate when there is actually 
a solution, its average of slow reads does not diverge from the average of 
slow reads in \sfw{}. This can be explained from the fact that read operations may be 
fast even when the predicate does not hold, but there is a confirmed tag propagated
in sufficient servers.  
%

\begin{figure*}[!ht]
{\bf 10 Readers:}\\	
	\begin{center}
	\vspace{-.7cm}
	\includegraphics[totalheight=2.3in]{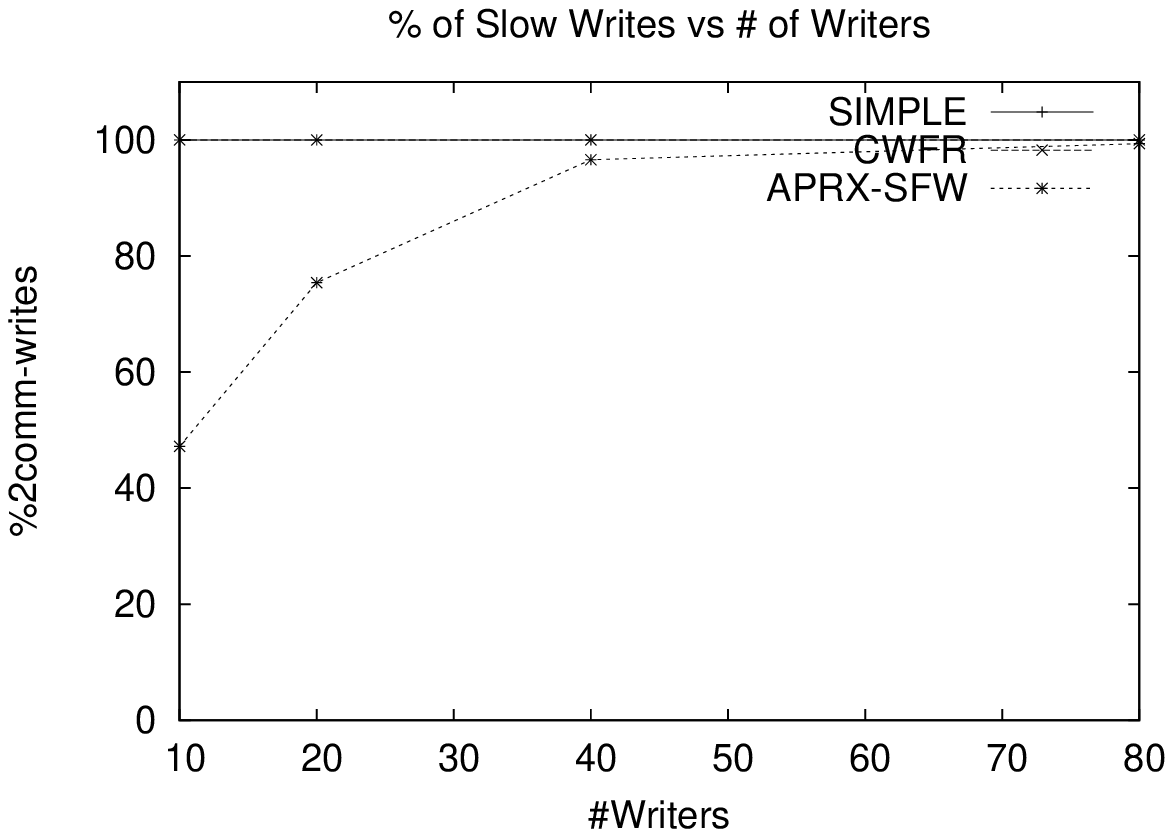}
	\hfill
	\includegraphics[totalheight=2.3in]{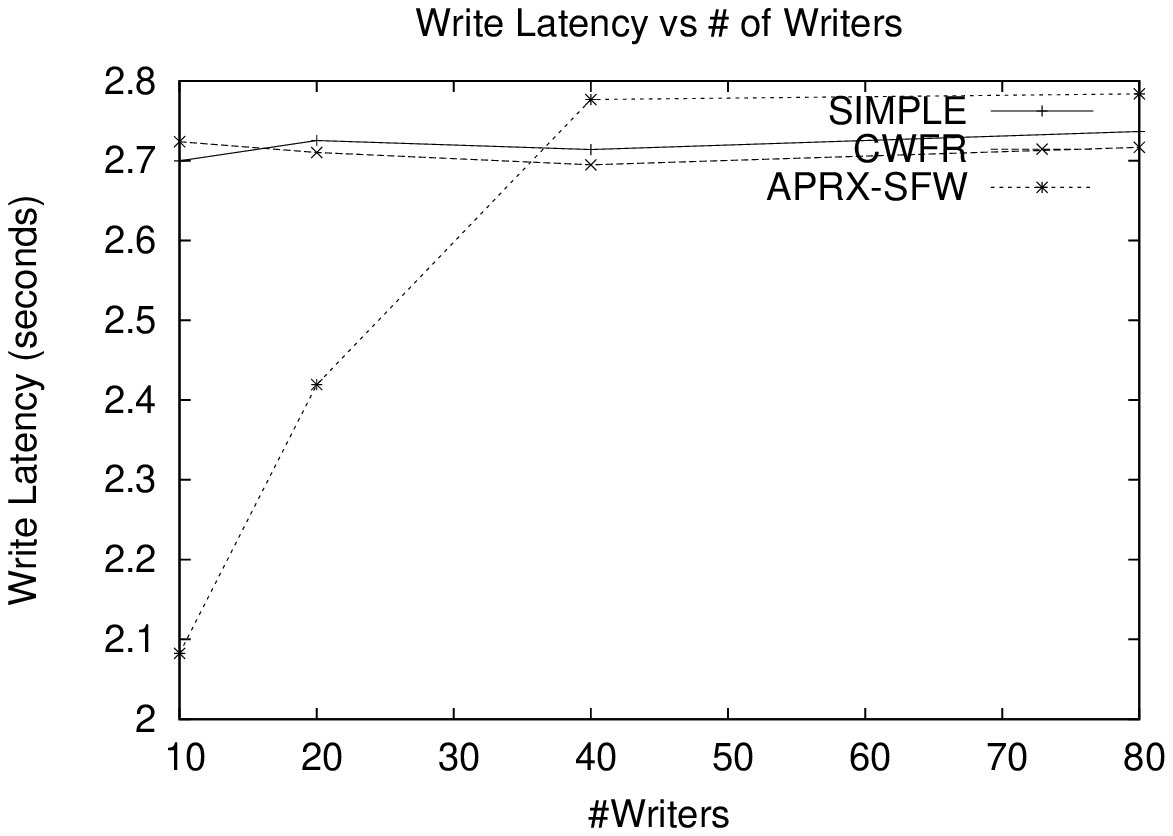}\\
	\end{center}
	\caption{14-wise quorum system ($|\srvSet=15$, $f=1$): {\bf Left Column:} Percentage of slow writes,  {\bf Right Column:} Latency of write operations }	
	\label{fig:sfwwrites}
\end{figure*}

A writer performs two rounds 
only when the predicate does not hold. Thus, counting the number of two-round writes
reveals how many times the predicate does not hold for an algorithm. 
According to our theoretical findings, algorithm 
\aprxsfw{} should allow no more than $\log|\srvSet| \cdot RR$ two-round reads or 
$\log|\srvSet| \cdot WR$ two-round writes 
in each scenario, where $RR$ and $WR$ are the number of two-round reads and writes 
allowed by the algorithm, respectively.
Our experimental results are within the theoretical upper bound, 
illustrating the fact that algorithm \aprxsfw{} implements a $\log|\srvSet|$-approximation 
relative to algorithm \sfw{}. Figure \ref{fig:sfwwrites} presents the average
amount of slow writes and the average write latency when we fix $|\rdSet|=10$,
$|\srvSet|=15$ and $f=1$. The number of writers vary from $|\wSet|=[10,20,40,80]$. 
This is one of the few scenarios we could run \sfw{} with 80 writers. As we can see 
the two algorithms experience a huge gap on the completion time of each write operation.
Surprisingly, \aprxsfw{} appears to win on the number of slow writes as well. Even though 
we would expect that \aprxsfw{} would contain more slow writes than \sfw{} this is not 
an accurate measure of the predicate validation. Notice that read operations may be invoked 
concurrently with write operations, and each read may also propagate a value in the system. 
This may favor or delay write operations. As the conditions on which the write operations 
try to evaluate their predicates are difficult to find, it suffices to observe that there
is a small gap on the number of rounds for each write for the two algorithms. 
Thus we claim the clear benefit of using algorithm \aprxsfw{} over 
algorithm \sfw{}.

\subsection{Algorithm \simple{} vs. \cwfr{} vs. \aprxsfw{}} 
\label{ssec:cwfr}

In this section we compare algorithm \aprxsfw{}
with algorithm \cwfr{}. To examine the impact of 
computation on the operation latency, 
we compare both algorithms to algorithm \simple{}. 
Recall that algorithm \simple{} requires insignificant 
computation. Thus, the latency of an operation in \simple{} 
directly reflects four communication delays (i.e., two rounds).

In the next paragraphs we present how the read and write operation 
latency is affected by the scenarios we discussed in Section \ref{ssec:ns2scenarios}.
A general conclusion that can be extracted from the simulations is that 
in most of the tests algorithms \aprxsfw{} and \cwfr{} perform better than 
algorithm \simple{}. This suggests that the additional computation 
incurred in these two algorithms does not exceed the delay
associated with a second communication round. 

\begin{figure*}[!ht]
{\bf 10 Writers:}\\	
	\begin{center}
	\vspace{-.7cm}
	\includegraphics[totalheight=2.3in]{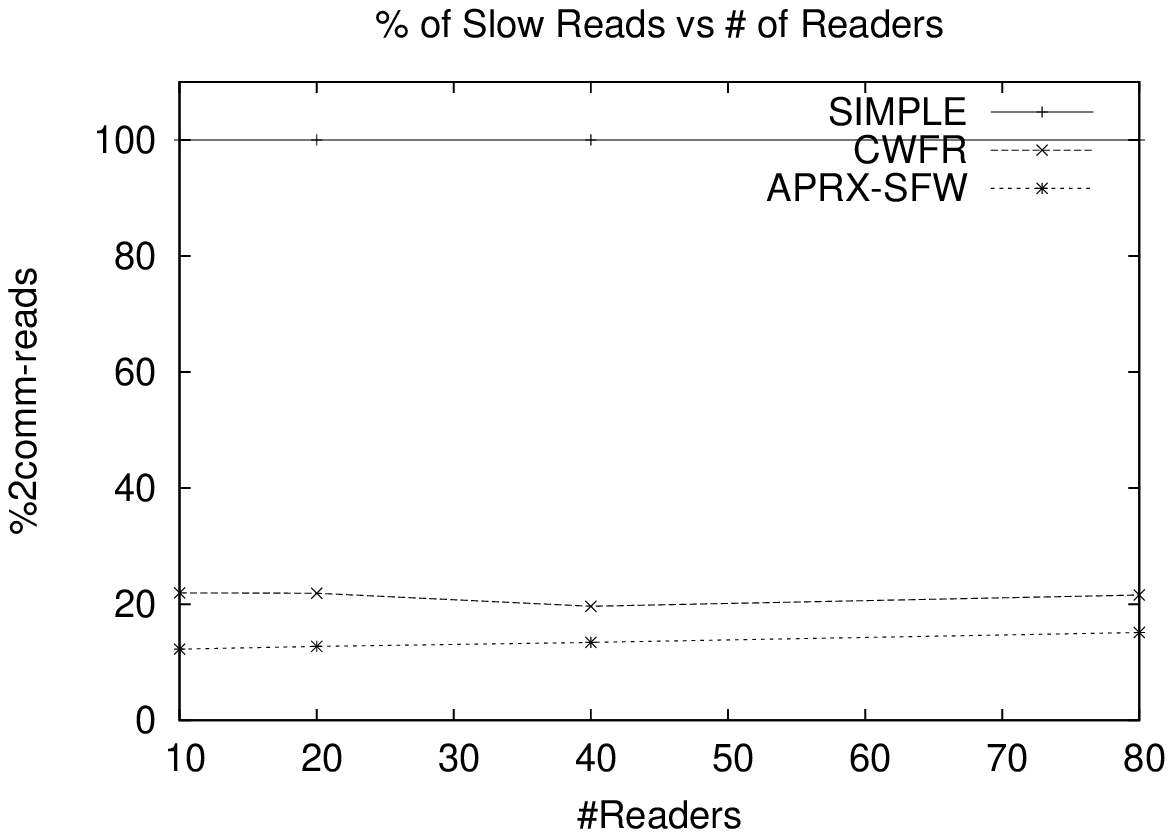}
	\hfill
	\includegraphics[totalheight=2.3in]{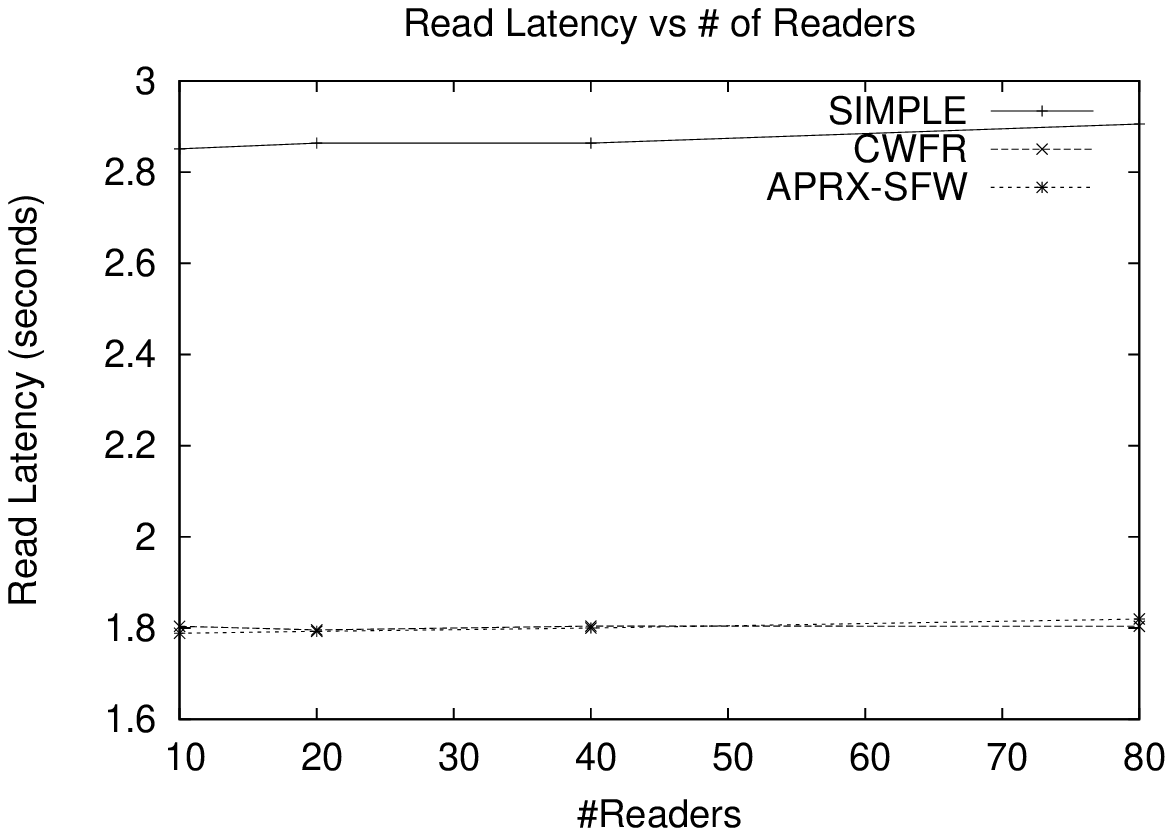}\\
	\end{center}
{\bf 80 Writers:}\\	
	\begin{center}
	\vspace{-.7cm}
	\includegraphics[totalheight=2.3in]{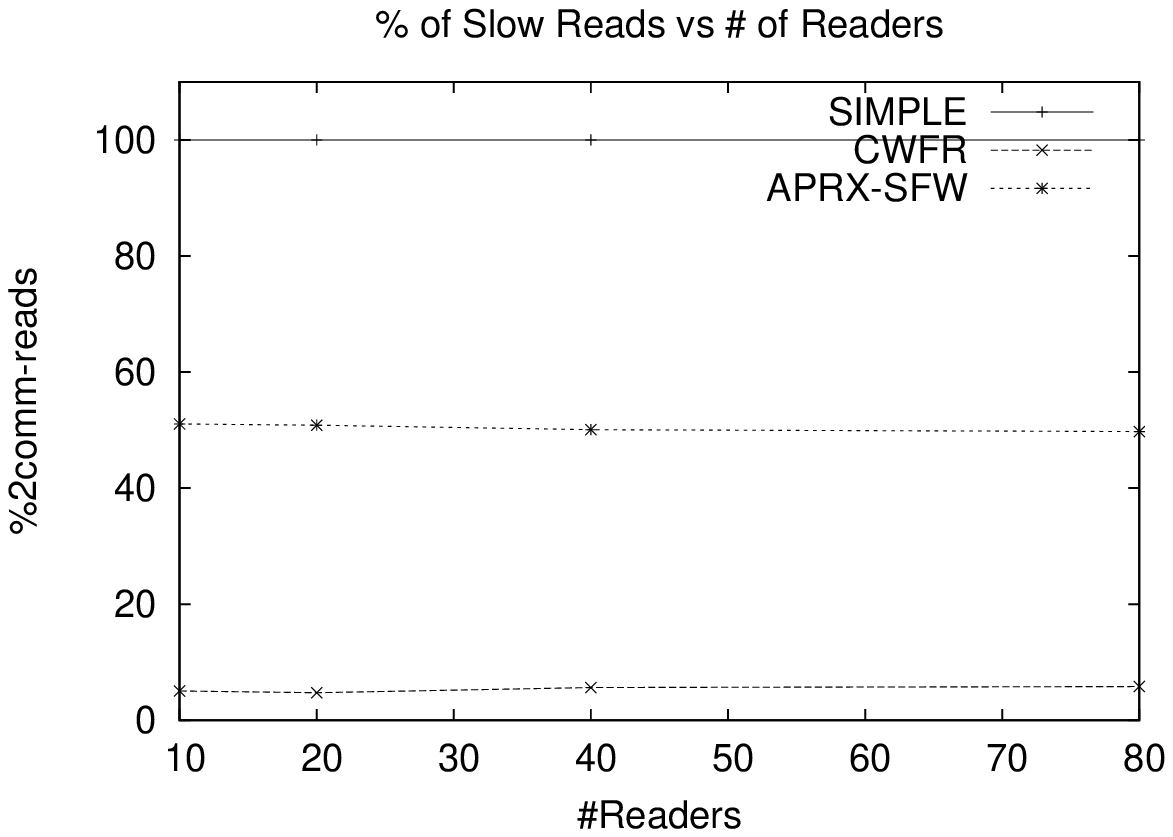}
	\hfill
	\includegraphics[totalheight=2.3in]{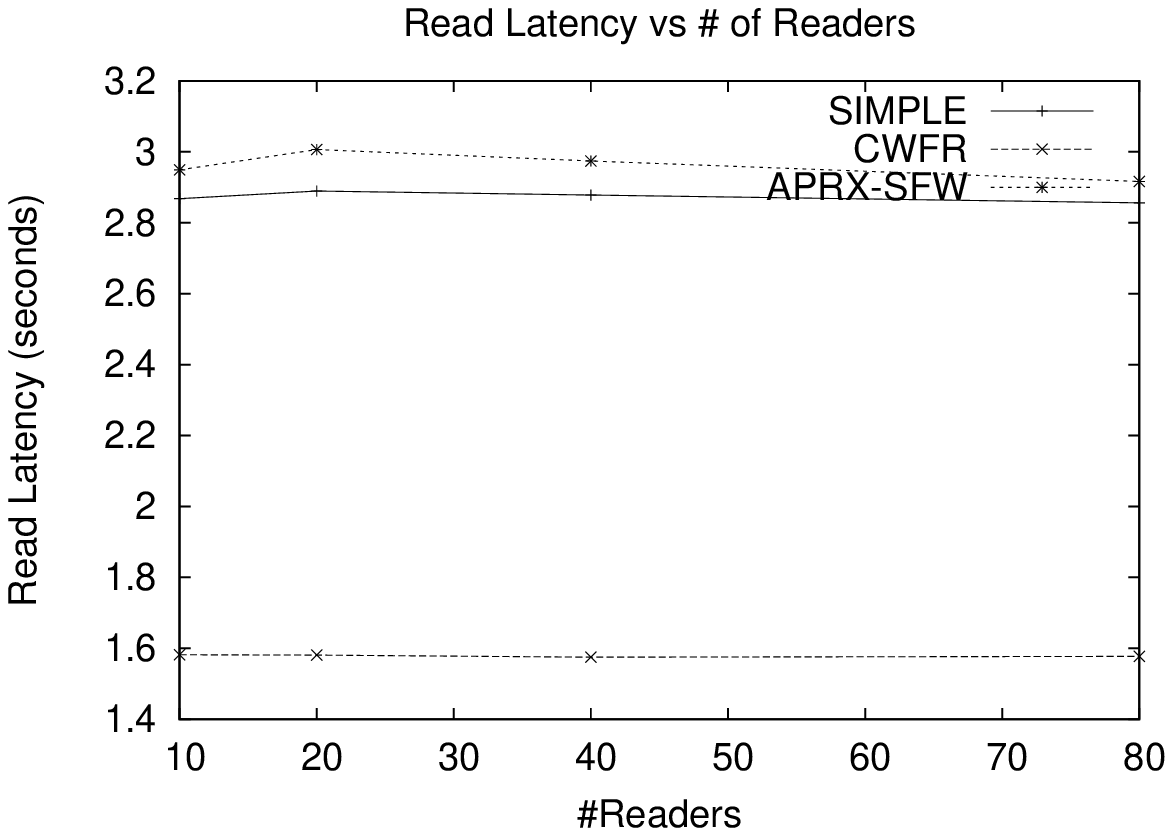}\\
	\end{center}
	\caption{19-wise quorum system ($|\srvSet|=20$, $f=1$): {\bf Left Column:} Percentage of slow reads,  {\bf Right Column:} Latency of read operations }	
	\label{fig:ns2readex}

\end{figure*}

\begin{figure*}[!ht]
{\bf 20 Readers:}\\	
	\begin{center}
	\vspace{-.7cm}
	\includegraphics[totalheight=2.3in]{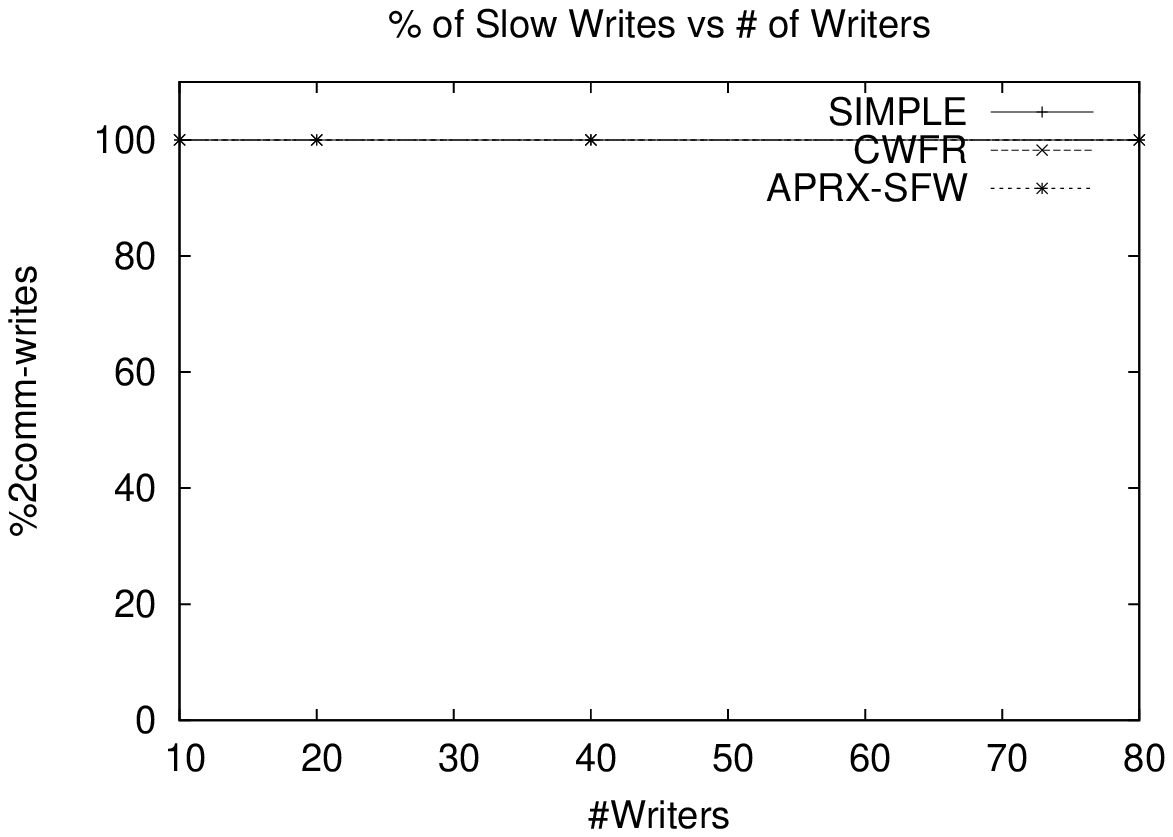}
	\hfill
	\includegraphics[totalheight=2.3in]{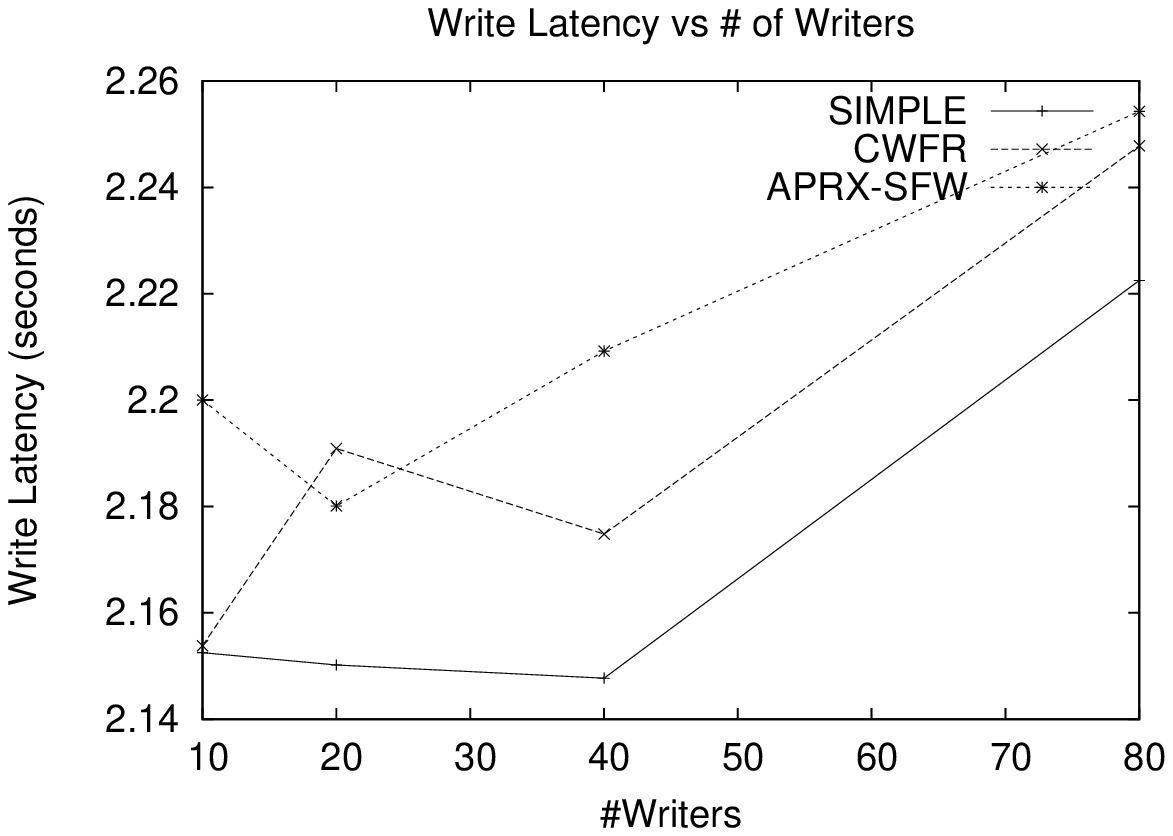}\\
	\end{center}
	\caption{4-wise quorum system ($|\srvSet=10$, $f=2$): {\bf Left Column:} Percentage of slow writes,  
	{\bf Right Column:} Latency of write operations}	
	\label{fig:ns2writeex}
\end{figure*}

\paragraph{Variable Participation:}
For this scenario we tested the scalability of the algorithms 
when the number of readers, writers, and servers changes. 
The plots that appear in Figures \ref{fig:ns2readex} and \ref{fig:ns2writeex}
present a sample of plots of this scenario for the read and write performance respectively.

Each figure contains the plots we obtain when we fix the number of 
servers and server failures, and we vary the number of readers and writers. 
So, for instance, Figure \ref{fig:ns2readex} fixes the number of 
servers to $|\srvSet|=10$, and $f=1$. Each row of the figure fixes the 
number of writers and varies the number of readers. Therefore we obtain four 
rows corresponding to $|\wSet|\in[10,20,40,80]$. 
Each row in a figure contains a pair of plots that presents the percentage 
 of slow reads (left plot) and the latency of each read (right plot) 
as we increase the number of readers. 

From the plots we observe that the read performance of neither algorithm 
is affected a lot by the number of readers in the system. On the other hand we observe
that the number of writers seem to have an impact on the performance of the 
\aprxsfw{} algorithm. As the number of writers grows we observe that 
both the number of slow read operations and the latency of read operations for 
\aprxsfw{} increases. 
Both \cwfr{} and \aprxsfw{} require 
fewer than 20\% of reads to be slow when no more than 20 writers exist in the service.
That is true for most of the server participation scenarios and leaves the latency of read
operations for the two algorithms below the latency of read operations in \simple{}. That 
suggests that the computation burden does not exceed the latency added by a second communication 
round. Once the number of writers grows larger than 40 the read performance of \aprxsfw{} 
degrades both in terms the number of slow reads and the average latency of read operations.
Unlike \aprxsfw, algorithm \cwfr{} is not affected by the participation of the 
service. The number of slow reads seem to be sustained below 30\% in all scenarios
and the average latency of each read remains under the 2 second marking.

\aprxsfw{} over-performs \cwfr{} only when the intersection degree is large and
the number of writers is small. The 20-wise (Figure \ref{fig:ns2readex})
intersection allows \aprxsfw{} to enable more single round read operations.
Due to computation burden however the average latency of each read in \aprxsfw{}
is almost identical to the average read latency in \cwfr{}. It is also evident
that the reads in \aprxsfw{} are greatly affected by the number of writers in the 
system. That was expected as by \aprxsfw{}, each read operation may examine 
a single tag per writer. From the plots on the second row of Figure \ref{fig:ns2readex}
we observe that although close to half reads are fast the average read latency of
\aprxsfw{} exceeds the average latency of \simple{}, or otherwise the average 
latency of 2 communication rounds. This is evidence that the computation time 
of \aprxsfw{} exceeded by a great margin the time required for a communication round.  

Similar to the read operations, the performance of write operations is not affected 
by the number of reader participants. It is affected however by both the number of
writers and servers in the system. The only algorithm that allows single round 
write operations is \aprxsfw{}. The number of writers and servers however, introduce
high computation demands for the write operations. As a result,
despite the fast writes, the average write latency of \aprxsfw{} can be higher than 
the average write latency of the other two algorithms. Note here that unlike the 
read operations, writes can be fast in \aprxsfw{} only when the write predicate holds. 
This characteristic can be also depicted from the plot presented in Figure \ref{fig:ns2writeex}
 where the 4-wise intersection does not allow 
for the write predicate to hold. Thus, every write operation in \aprxsfw{} performs
two communication rounds in this case. The spikes on the latency of the write operations 
in the same figure appear due to the small range of the values and the small time inconsistency 
that may be caused by the simulation randomness.

\paragraph{Quorum Construction:}
We consider majority quorums due to the property 
they offer on their intersection degree \cite{EGMNS09}:
if $|\srvSet|$ the number of servers and up to $f$ of them may crash
then if every quorum has size $|\srvSet|-f$ we can construct
a quorum system with intersection degree $n=\frac{|\srvSet|}{f}-1$.
Using that property we obtain the quorum systems presented on Table \ref{tab:quorums} 
by modifying the number of servers and the maximum number of server failures.

\begin{table}
	\begin{center}
		\begin{tabular}{|c|c|c|c|}
		\hline
		{\bf Servers} & {\bf Server Failures} & {\bf Int. Degree} & {\bf Quorums}  \\
		 $|\srvSet|$ &  $f$ & $n$ & $|\qs|$ \\
		\hline
		10 & 1 & 9 & 10 \\
		\hline
		15 & 1 & 14 & 15 \\
		\hline
		20 & 1 & 19 & 20 \\
		\hline
		25 & 1 & 24 & 25 \\
		\hline
		10 & 2 & 4 & 45 \\
		\hline
		15 & 2 & 6 & 105 \\
		\hline
		20 & 2 & 9 & 190 \\
		\hline
		25 & 2 & 11 & 300 \\
		\hline
		\end{tabular}
	\end{center}
	\caption{Quorum system parameters.}
	\label{tab:quorums}
\end{table}

\begin{figure*}[!ht]
{\bf 40 Readers, 80 Writers, f=2:}\\	
	\begin{center}
	\vspace{-.7cm}
	\includegraphics[totalheight=2.3in]{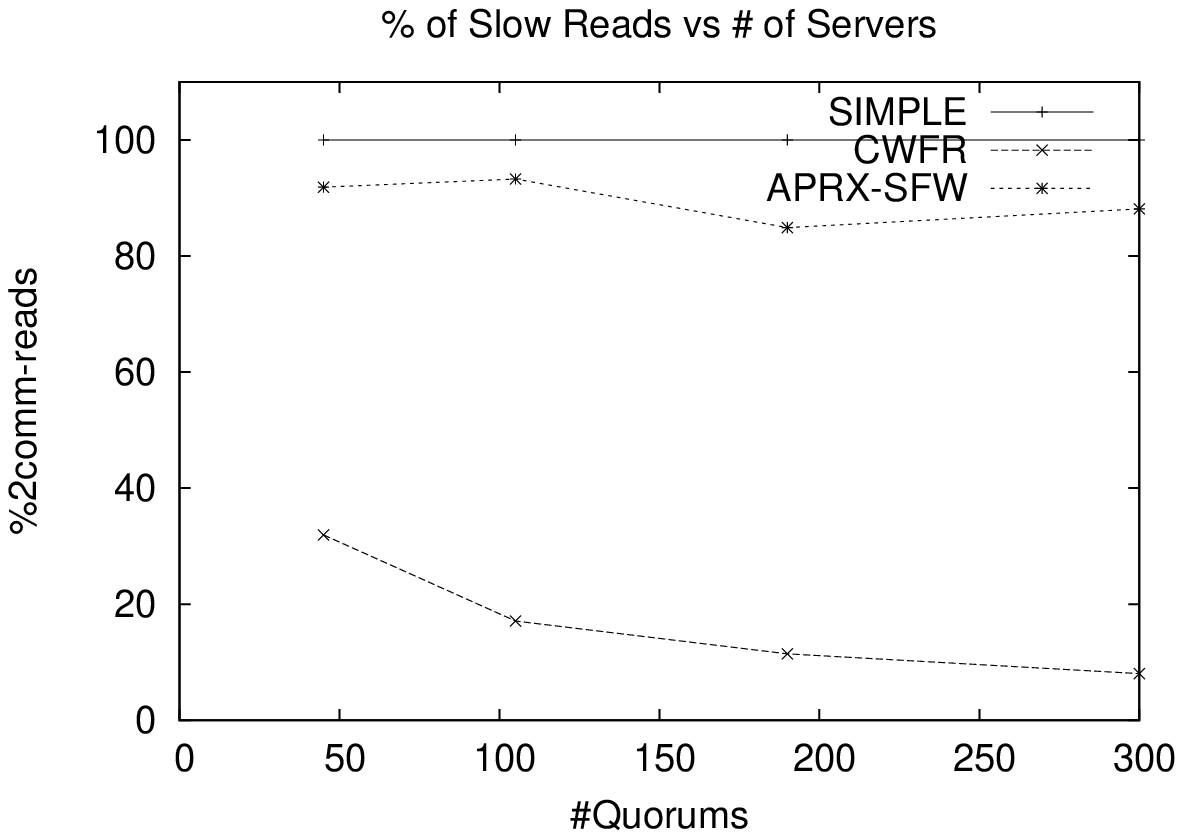}
	\hfill
	\includegraphics[totalheight=2.3in]{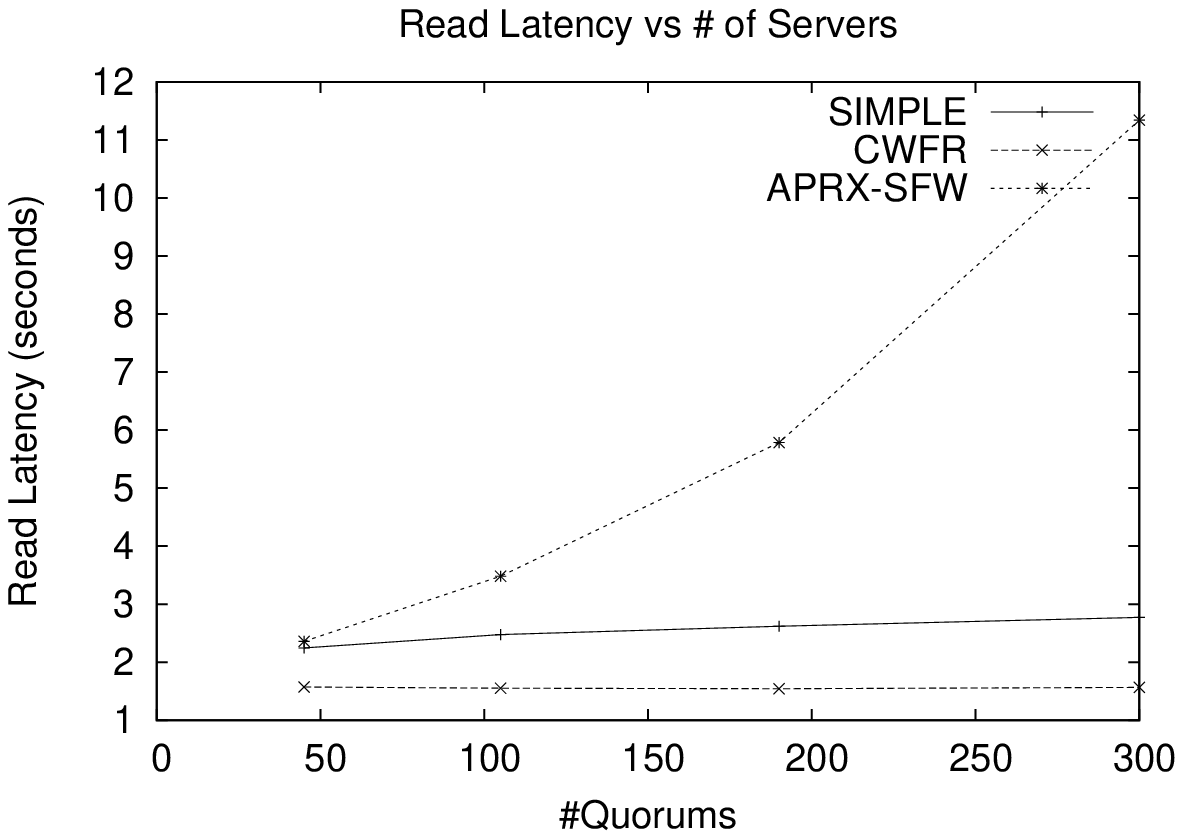}\\
	\includegraphics[totalheight=2.3in]{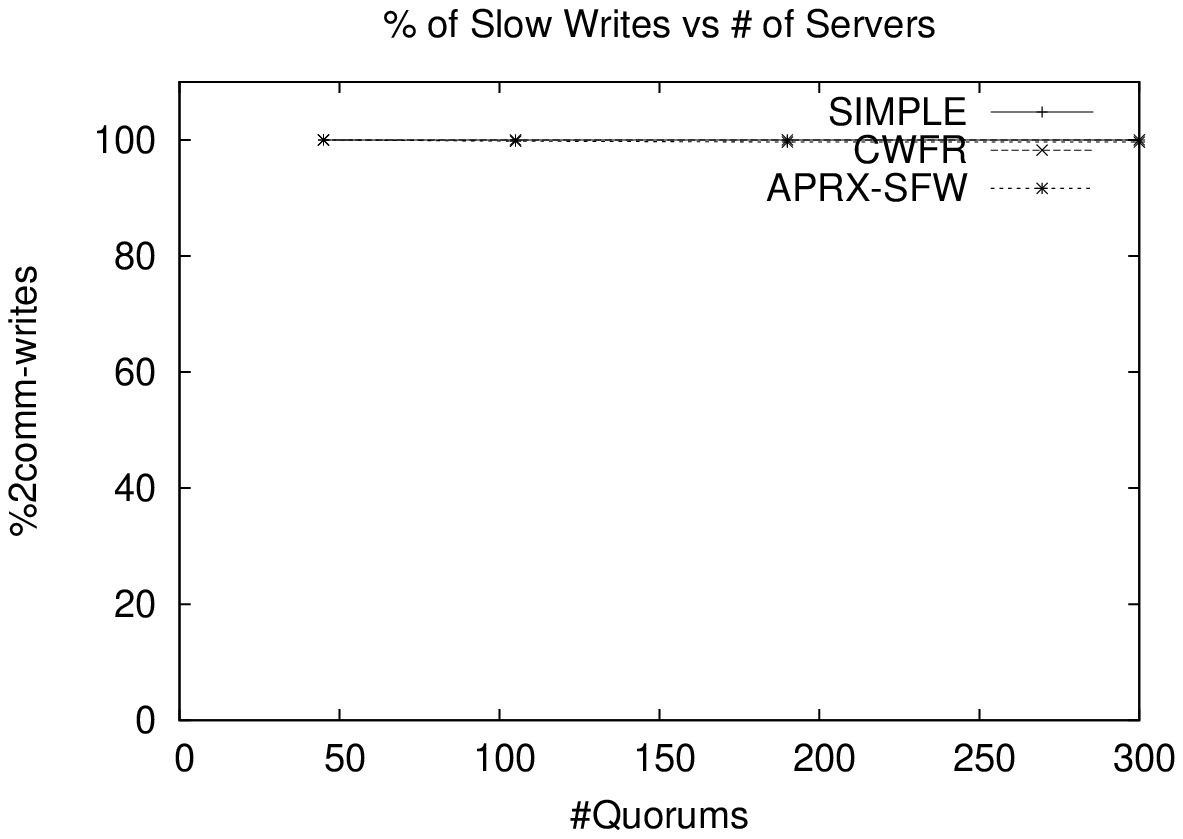}
	\hfill
	\includegraphics[totalheight=2.3in]{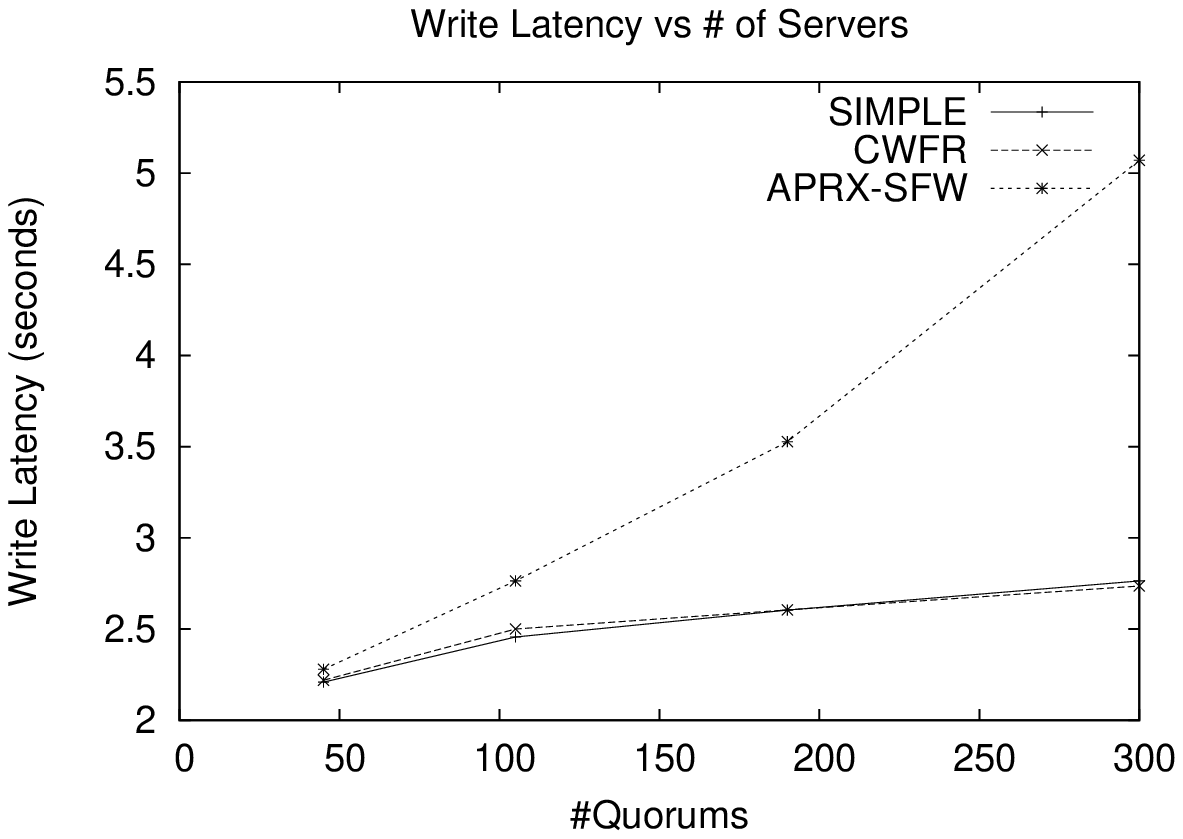}\\
	\end{center}
	\caption{{\bf Left Column:} Percentage of slow operations,  {\bf Right Column:} Latency of operations}	
	\label{fig:ns2quorumsex}
\end{figure*}

Figure \ref{fig:ns2quorumsex} plots the performance of 
read and write operations (communication rounds and latency) with 
respect to the number of quorum members in the quorum system.
The figure two pairs that  correspond to the quorum system that allows 
up to two server failures. The top pair
describes the performance of read operations while the bottom pair
the performance of write operations. Lastly, the left plot in each pair 
presents the percentage of slow read/write operations, and the right
plot the latency of each operation respectively.

We observe that the incrementing number of servers,
and thus cardinality of the quorum system, reduces the percentage of slow reads for 
both \aprxsfw{} and \cwfr{}. Operation latency on the other hand is not proportional to 
the reduction on the amount of slow operations. Both algorithms 
\aprxsfw{} and \cwfr{} experience an incrementing trend on the latency of 
the read operations as the number of servers and the quorum members
increases. Worth noting is that the latency of read operations in \simple{} 
also follows an increasing trend even though every read operation requires
two communication rounds to complete. This is an evidence that the increase on the 
latency is partially caused by the communication between the readers and the servers:
as the servers increase in number the readers need to send and wait for more messages.
The latency of read operations in \cwfr{} is not affected greatly as the 
number of servers changes. 
As a result, \cwfr{} appears to maintain a read latency close to 1.5 sec in 
every scenario. With this read latency \cwfr{} over-performs algorithm \simple{}
in every scenario as the latter maintains a read latency between 2.5 and 3 sec.
Unlike \cwfr{}, algorithm \aprxsfw{} experiences a more aggressive change 
on the latency of read operations. The read latency in \aprxsfw{} is affected 
by both the number of quorums in the system, and the number of writers in the system. 
As a result the latency of read operations in \aprxsfw{} in conditions with 
a large number of quorum members and writers may exceed the read latency of 
\simple{} up to 5 times. The reason for such performance is that every 
read operation in \aprxsfw{} the reader examines the tags assigned to 
every writer and for every tag runs the approximation algorithm on a number of 
quorums in the system. The more the writers and the quorums in the system, the more
time the read operation takes to complete. 

Similar observations can be made for the write operations. 
Observe that although both \cwfr{} and \simple{} require 
two communication rounds per write operation, the write 
latency in these algorithms is affected negatively by the 
increase of the number of quorums in the system. As we said before
this is evidence of the higher communication demands when we increase
the number of servers. We also note that the latency of the write operations
in \simple{} is almost identical to the latency of the read operations of
the same algorithm. This proves the fact that the computation demands in either 
operation is also identical. As for \aprxsfw{}, the increase 
on the number of servers reduces the amount of slow write operations.
The reduction on the amount of the slow writes is not proportional
to the latency of each write. Thus, the average write latency of \aprxsfw{} 
increases as the number of quorums increases in the system. 
Unlike the latency of read operations, the latency of writes do not
exceed the write latency of \simple{} by more than 3 times. Comparing with 
the latency of read operations it appears that although \aprxsfw{} may allow 
more fast read operations than writes under the same conditions, 
the average latency of each read is higher than the latency of write operations.
An example of this behavior can be seen in Figure \ref{fig:ns2quorumsex}.
In this example less than 90\% of reads need to be slow and the average read latency
climbs to almost 12 seconds. On the other hand almost every write operation is slow
and the average write latency climbs just above 5 seconds. The simple explanation for 
this behavior lie on the evaluation of the read and write predicates. Each reader needs
to examine the latest tags assigned to every writer in the system whereas each writer only 
examines the tags assigned to its own write operation.


\paragraph{Network Latency:}
During our last scenario we considered increasing the latency of the 
network infrastructure from 10ms to 500ms. With this scenario we want to examine whether in slow
networks is more preferable to minimize the amount of 
rounds, even if that means higher computation demands. 
We considered just a single setting for this scenario where the 
number of servers is 15, the maximum number of failures is 2 and 
we pick the number of readers and writers to be one of [10, 20, 40, 80].

\begin{figure*}[!ht]
{\bf 20 Writers:}\\	
	\begin{center}
	\vspace{-.7cm}
	\includegraphics[totalheight=2.3in]{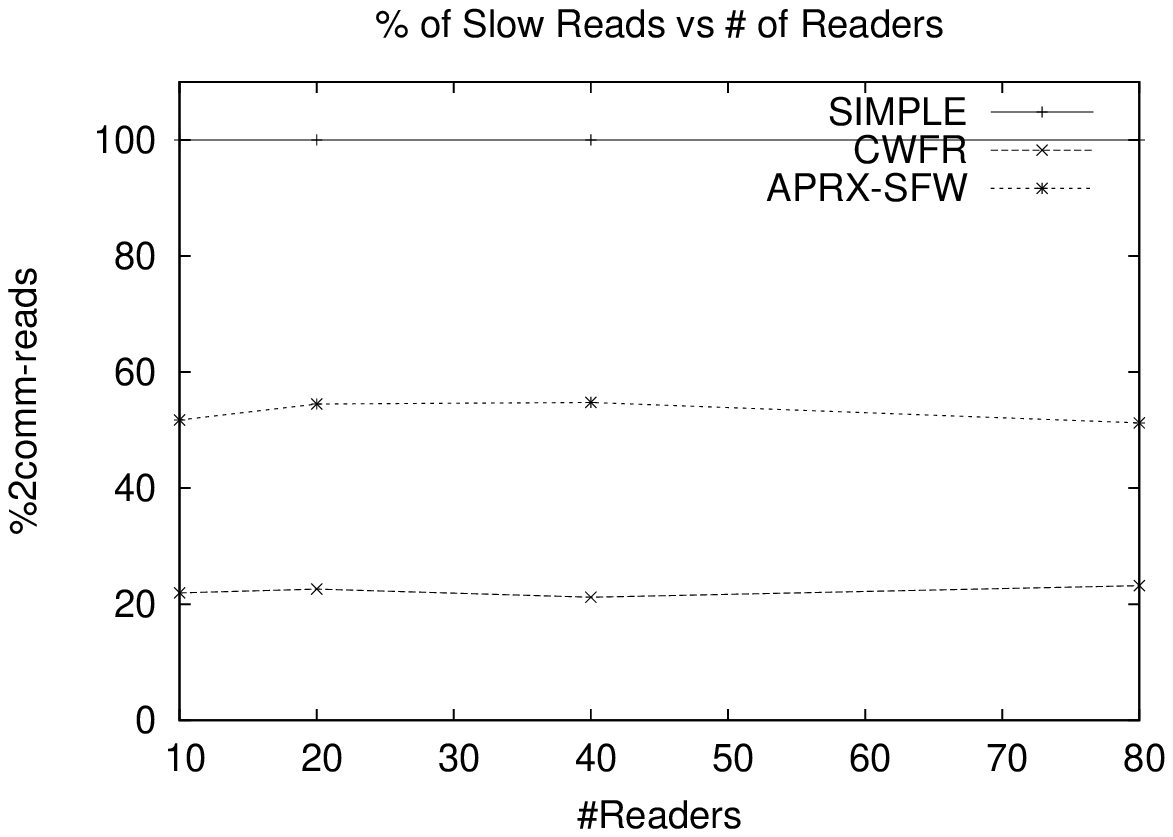}
	\hfill
	\includegraphics[totalheight=2.3in]{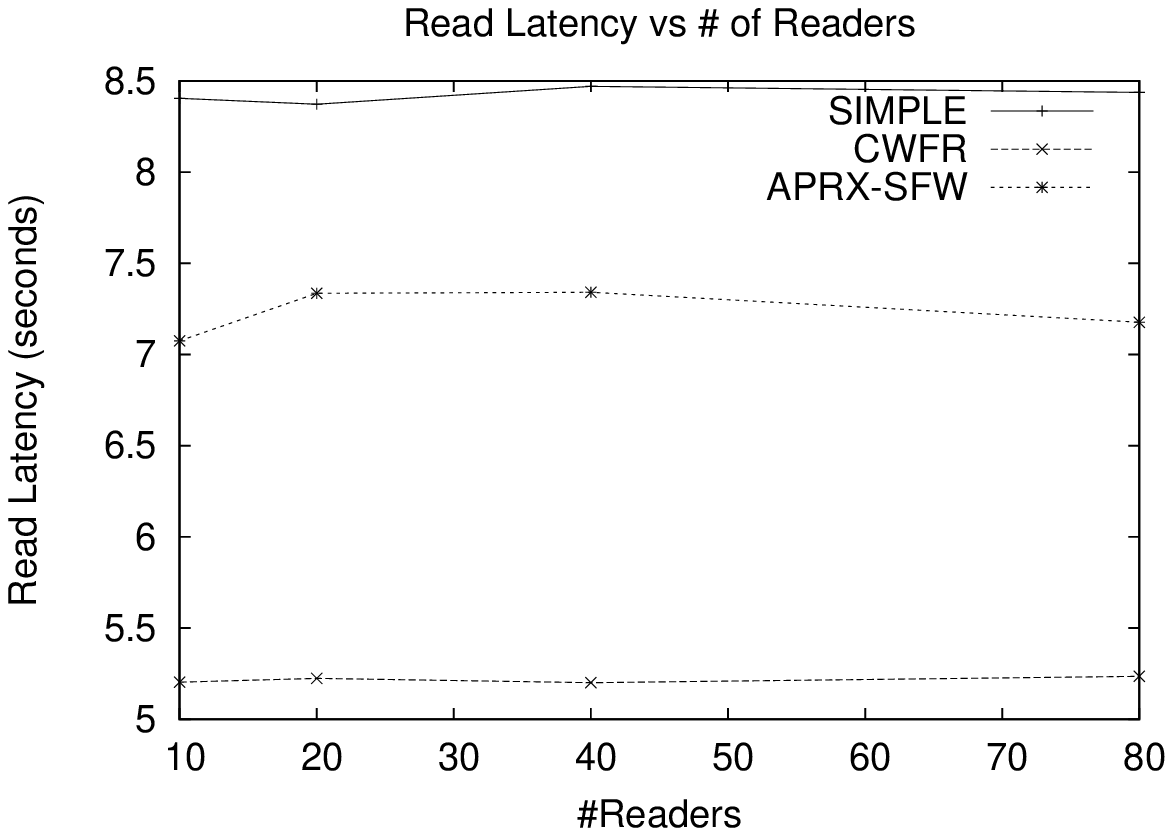}\\
	\includegraphics[totalheight=2.3in]{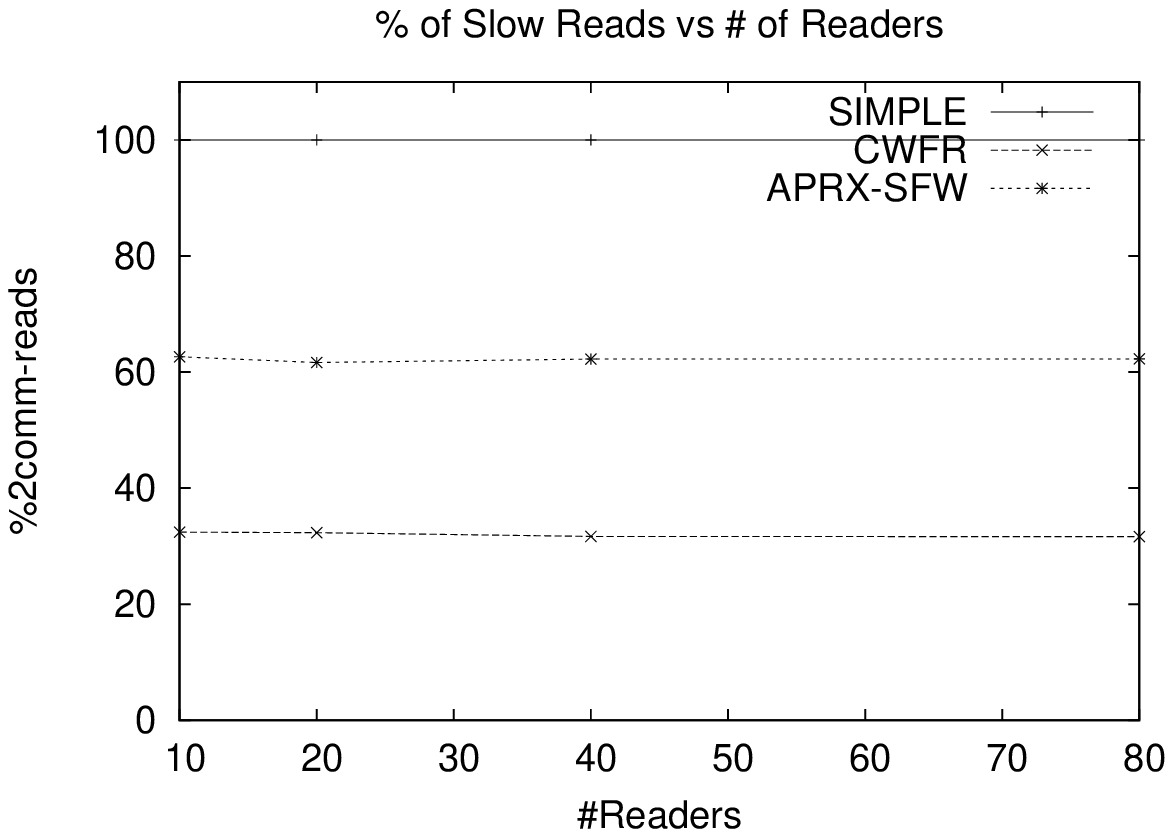}
	\hfill
	\includegraphics[totalheight=2.3in]{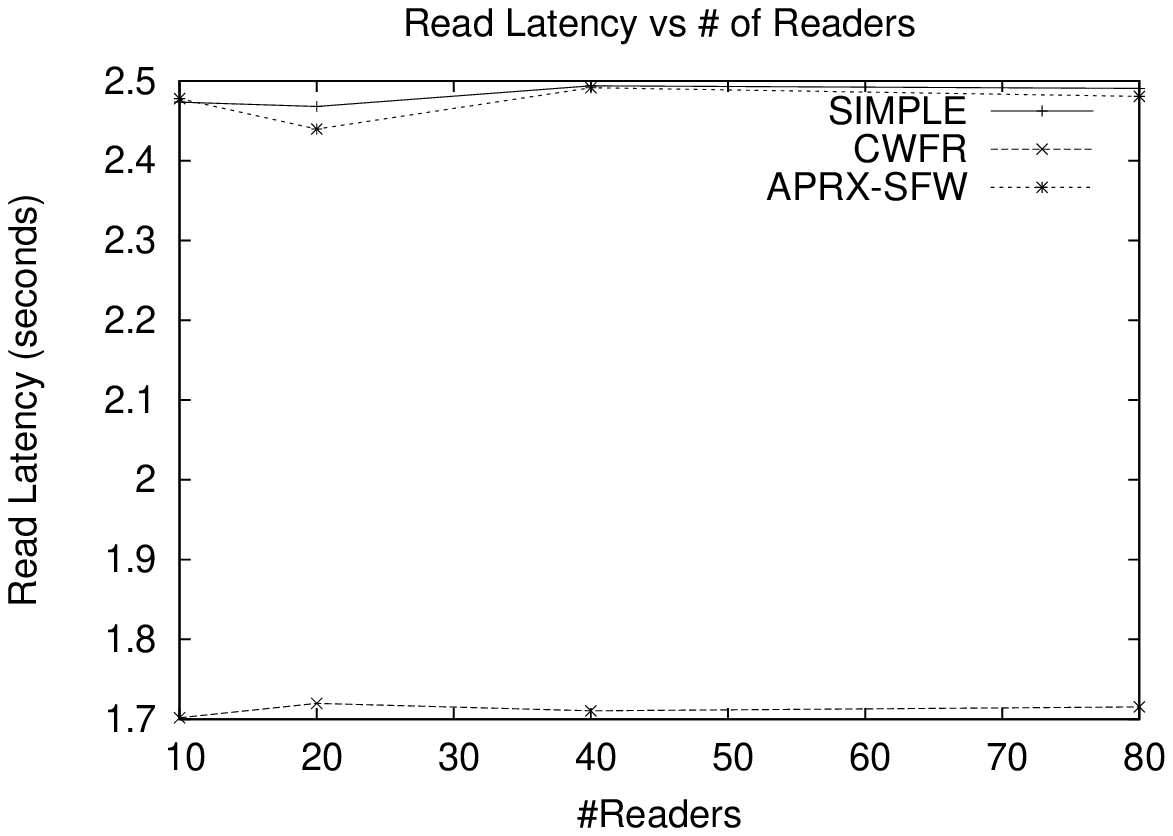}\\
	\end{center}
	\caption{6-wise quorum system ($|\srvSet|=15$, $f=2$): {\bf Left Column:} Percentage of slow reads,  {\bf Right Column:} Latency of read operations}	
	\label{fig:ns2latencyex}
\end{figure*}

In order to establish meaningful conclusions we need to compare the 
outcomes of the operation performance of this scenario with the 
respective scenario where the latency is 10ms. We notice that the 
largest network delay reduces the amount of slow reads for both \cwfr{} and 
\aprxsfw{}. In addition the delay indeed helps \aprxsfw{} to perform better 
than \simple{} in scenarios where \aprxsfw{} was performing identical or worse 
than \simple{} when the delay was 10ms. 
This can be seen in Figure \ref{fig:ns2latencyex}. As we can see in the figure the latency of the
read operations of \aprxsfw{} was aligning with the read latency of \simple{} 
when the delay was 10ms. When we increased the network delay to 500ms the 
average read latency of \aprxsfw{} was remarkably smaller than the average 
latency of \simple{} under the same participation and failure conditions.
Similar observations can be made for the write operations. 

So we can safely conclude that the network delay can be one of the factors 
that may affect the decision on which algorithm is suitable for a particular 
application.

\section{PlanetLab Implementation}
\label{sec:pl}
In this section we present our experiment on the Planetlab platform.
First we provide a description of the PlanetLab platform and then 
we present the parameters we considered and 
the scenarios we run for our implementations. 

\subsection{The PlanetLab Platform}
\label{ssec:pl}
PlanetLab is an overlay network infrastructure which is available as a testbed for 
computer networking and distributed systems research. As of September 2011, PlanetLab is 
composed of 1075 machines, provided by academic and industry 
institutions, at 525 locations worldwide. 
Malicious and buggy services can affect the communication infrastructure and the 
nodes' performance; therefore, strict terms and conditions for providing security and stability 
in the PlanetLab are enforced. Nodes may be installed or rebooted at any time, turning their disk 
into a temporary form of storage, providing no guarantee regarding their reliability.       

As oppose to NS2 simulator \cite{NS2}, Planetlab provides no control over the components and execution sequence 
of the algorithms, as it overheres the communication delays and congestion conditions 
of wide area networks (e.g. Internet cloud). Hence, some of the execution scenarios and environmental settings used for the simulation environment do not apply in this adverse and unpredictable, large-scale, real-time environment. Furthermore, since no global snapshot of the system may be acquired, we rely on local decisions and readings of each individual participant in the system. 

\subsection{Experimentation Platform}
\label{ssec:pltestbed}

Our test environment consists
of a set of writers, readers, and servers. 
Communication between the nodes is established via 
TCP/IP. For our experiments we used the machines that
appear in Table \ref{tab:machines}. Those were in total 20 machines.
Our implementations were written in C++ programming language
and C sockets were used for interfacing with TCP/IP.  

\begin{table}
\begin{center}
\begin{tabular}{|l|c|c|c|c|}
\hline
{\bf Machine name} & {\bf Country} & {\bf \# Cores} & {\bf CPU Rate (Ghz)} & {\bf RAM (GB)} \\
\hline
nis-planet2.doshisha.ac.jp & Japan & 2 & 2.9 & 4 \\
\hline
chronos.disy.inf.uni-konstanz.de & Germany & 2 & 2.4 & 4 \\
\hline
planetlab2.cs.unc.edu & United States & 4 & 2.6 & 4 \\
\hline
peeramidion.irisa.fr & France & 2 & 2.9 & 4 \\
\hline
pl2.eng.monash.edu.au & Australia & 4 & 2.6 & 4 \\
\hline
planetlab2.ceid.upatras.gr & Greece & 2 & 2.6 & 4 \\
\hline
planetlab-01.bu.edu & United States & 8 & 2.4 & 4 \\
\hline
planetlab3.informatik.uni-erlangen.de & Germany & n/a & n/a & n/a \\
\hline
planetlab1.cs.uit.no & Norway & 2 & 2.6 & 4 \\
\hline
ple2.cesnet.cz & Czech Republic & 4 & 2.8 & 8 \\ 
\hline
planetlab01.tkn.tu-berlin.de & Germany & 4 & 2.4 & 4 \\
\hline
planetlab1.unineuchatel.ch & Switzerland & 2 & 2.6 & 4 \\
\hline
evghu14.colbud.hu & Hungary & n/a & n/a & n/a  \\
\hline
zoi.di.uoa.gr & Greece & 4 & 2.4 & 8 \\
\hline
planetlab1.cs.vu.nl & Netherlands & 8 & 2.3 & 4 \\
\hline
pli1-pa-4.hpl.hp.com & United States & 8 & 2.5 & 6 \\
\hline
plab3.ple.silweb.pl & Poland & 4 & 2.4 & 4 \\
\hline
planetlab-2.imperial.ac.uk & United Kingdom & 2 & 2.9 & 1 \\
\hline
planet1.unipr.it &  Italy & 2 & 2.6 & 4 \\
\hline
planetlab2.rd.tut.fi & Finland & 8 & 2.4 & 16 \\
\hline
\end{tabular}
\end{center}
\caption{List of machines that hosted our experimet.} 
\label{tab:machines}
\end{table}

%
Each of those machines was used to host one or more processes. 
The servers were hosted on the first 10 machines. If the number of
servers exceeded the number of hosts then each host was running 
more than one server instance in a round robin fashion. 
For instance, if the number of servers were 15 then two 
servers were running in each of the first 5 machines, and one 
server instance on each of the remaining 5 machines. 
In other words the $1^{st}$ and $11^{th}$ servers were running on 
the first machine, the  $2^{nd}$ and $12^{th}$ servers on the second 
machine and so on. Note that we choose the first 10 
machines such as to split our servers geographically throughout the world. 
The readers and the writers were started on all 20 machines starting 
from the $10^{th}$ machine and following a round robin technique. 
Thus, some of the first 10 machines could run a server and a client 
at the same time. 
 
We have evaluated the algorithms with majority quorums. 
 As discussed in \cite{EGMNS09}, assuming $|\srvSet|$ servers
 out of which $f$ can crash, we can construct an
 $(\frac{|\srvSet|}{f}-1)$-wise quorum system $\qs$. 
 Each quorum $Q$ of $\qs$ has size $|Q|=|\srvSet|-f$. 
 The processes are not aware of $f$.  The quorum system is 
generated \emph{a priori} and is distributed to each machine 
as a file. So each participant can obtain the quorum construction
be reading the appropriate file from the host machine.


We use the positive time parameters $rInt=10sec$ and $wInt=10sec$
 to model operation frequency. Readers and writers pick a uniformly at random 
 time between $[0\ldots rInt]$ and $[0\ldots wInt]$, respectively,
 to invoke their next read (resp. write) operation. Each  reader and 
writer chooses a new timeout every time their latest operation is completed. 
This preserves the property of well-formedness discussed in Section~\ref{sec:model}.
%

Finally we specify the number of operations each participant should 
invoke. For our experiments we allow participants to perform up to 20 
operations (this totals to 400-3000 operations). 

\subsection{Difficulties}
\label{ssec:pldif}
Consistent storage protocols are not readily designed for the network 
infrastructure. For this reason our algorithms need to utilize the 
existing communication protocols like TCP/IP to achieve process communication. 
So we need to be especially careful so that the use of these protocols does 
not affect the correctness of our algorithms. Below we present some 
hurdles we faced during the development of our implementations along 
with the workarounds we used to ensure the correctness of the algorithms.

\paragraph{Multiplexing.}
Client-Server architecture proved to be more challenging than expected 
for the implementations of consistent implementations. In traditional techniques 
the server listens for an incoming connection and spawns a child or a thread process 
to handle an incoming request from a client. 
By doing so, each client is being served independently from the rest of the 
clients, and thus does not have to wait for the other clients to terminate. 
At the same time each client gets the illusion that is the only one 
communicating with the server. 
This is acceptable if clients do not modify the local information of the 
server. It creates serious synchronization issues however when the 
two clients try to update the server's state.

To overcome this problem we chose not to use forking or threads to establish the 
communication between the server with the clients. Rather, we used \emph{multiplexing}.
This is a non-blocking technique where the server does not block to wait 
for incoming connections. Instead, the server allows the clients to connect to it 
and periodically checks of any new connection request. Once a new connection is 
established the server generates a new file descriptor and places the descriptor
in a pool of connections. In a continuous loop the server checks if a client wants
to transmit a message by checking the state of the descriptor associated with the 
connection of the particular client. When a message is received the server communicates
with the sending process to satisfy its request. This is an explicit communication 
between the server and the client and circumvents any synchronization issues. On the
other hand, the use of non-blocking communication protocols allows different clients 
to be connected at the same time with a particular server and wait their turn until they
get served. 


\paragraph{Resource Limits.}
Each slice on PlanetLab offers limited resources for each experiment. 
Furthermore the master machine that starts the experiments has limitations 
on the number of processes that can be initiated concurrently. 
For these reasons we bounded the number of readers and writers to 40.
We still obtain four points by running our implementations with 
 10,20,30 and 40 read and write processes. We believe that this is 
an adequately large sample for the extraction of consistent results. 

\paragraph{Sampling.}
The PlanetLab environment is extremely adverse. 
Machines may go down at any time in the execution, the network 
latency may increase without notice and participant communication 
may be interrupted. Thus, to obtain some reasonable results we had 
to run our algorithms alternatively to maintain short time intervals 
between the algorithm's runs. 
Our first try was to run all the scenarios on a single algorithm
and then move on to the next algorithm. This technique produced large 
time intervals between the execution of two consecutive algorithms.
The problem was that the conditions (network delay, node failures etc.) 
of the network at the time of the run of 
one algorithm were changing dramatically by the time the next algorithm 
was running. Algorithm alternation proved to provide a solution to this
problem. Even though we still observe network changes the plotting of the 
results provide less ``noise'' on the latency of the operations due to 
network delay differences.  

\paragraph{Weighted Average Time.}
The last issue we faced during our development had to do with the 
hectic participation of the nodes in PlanetLab. Since the nodes in 
PlanetLab may be interrupted at any time during the execution 
of any algorithm, we observed that in most of the runs not all the 
readers were terminating. This is because the readers or writers hosted in a failed
PlanetLab node were not able to complete all the necessary operations. 
As the operations completed by those failed readers and writers were 
not counted, getting the average of the read/write latency by dividing 
the total latency over the total number of terminated readers/writers per run did not 
produce a consistent result. So we decided to obtain the weighted average
of the read/write latency by dividing the total operation latency over the 
total number of terminated processes. For example, assume a scenario were 
we initiated 20 readers out of which 10 terminated in the first 
run and 15 terminated in the second run. Lets assume for example that the total read latency
of the first run was 100s and the total latency of the second run was 120s. With the 
non-weigthed method the average latency was 
\[
	nonWeightedAvg= (\frac{100}{10}+\frac{120}{15})/2 = 9s
\] 
With the weigthed average we obtain the following:
\[
	WeightedAvg= \frac{100+120}{10+15} = 8s
\]
The weighted average is the average latency we use for our plots.

\subsection{Scenarios}
\label{ssec:plscenarios}

The scenarios were designed to test
(i) scalability of the algorithms as the number of 
readers, writers and servers increases, and (ii) the relation between 
 quorum system deployment and operation latency. 
In particular we consider the following parameters:
\begin{enumerate}

	\item {\bf Number of Participants:} We run every test with 10, 20, 30, and 40 readers and writers.
	To test the scalability of the algorithms with respect to the number
	of replicas in the system we run all of the above tests with 10, 15, 20, and 25 servers. 
	Such tests highlight whether an algorithm is affected by the number of participants in the system. 
	Changing the number of readers and writers help us investigate how each algorithm 
	handles an increasing number of concurrent read and write operations. The more the servers
	on the other hand, the more concurrent values may coexist in the service. So, algorithms 
	like \aprxsfw{} and \cwfr{}, who examine all the discovered values and do not rely on the 
	maximum value, may suffer from local computation delays.

	\item {\bf Quorum System Construction:} As in the NS2 simulation we use
	majority quorums as they can provide quorum systems with high intersection degree.
	So, assuming that $f$ servers may crash we construct quorums of size 
	$|\srvSet|-f$. 
	As the number of servers $|\srvSet|$ varies between 10,15,20, and 25, we run the tests for 
	two different failure values, i.e. $f=1$ and $f=2$. This affects our environment in two ways:
	\begin{itemize}
		\item[(i)] We get quorum systems with different quorum intersection degrees. 
		According to \cite{EGMNS09} for every given $\srvSet$ and $f$ we obtain a $(\frac{|\srvSet|}{f}-1)$-wise quorum system.
		\item[(ii)] We obtain quorum systems with different number of quorum members. For example assuming
		15 servers and 1 failure we construct 15 quorums, whereas assuming 15 servers and 2 failures 
		we construct 105 different quorums. 
	\end{itemize}
	Changes on the quorum constructions help us evaluate how the algorithms
	handle various intersection degrees and quorum systems of various memberships. 
	%

\end{enumerate}

Another parameter we considered was operation frequency. Due to the invocation of operations 
in random times between the read and write intervals as explained in Section \ref{ssec:pltestbed},
operation frequency varies between each and every participant. Thus, fixing different initial 
operation frequencies does not have an impact of the overall performance of the algorithms. For this
reason we avoided running our experiments over different operation frequencies.

\section{Planetlab Results}
\label{sec:plres}
 
In this section we discuss our findings. 
We compare algorithms \cwfr{}, \aprxsfw{}, and 
\simple{} to establish conclusions on the overall performance 
(including computation and communication) of the algorithms. 
All the plots obtained by this experiment appear in \cite{D6-TR}. 

To examine the impact of computation on the operation latency, 
we compare the performance of algorithms~\aprxsfw{}
and~\cwfr{} 
to the performance of algorithm~\simple{}. 
Recall that algorithm \simple{} requires insignificant 
computation. Thus, the latency of an operation in \simple{} 
directly reflects four communication delays (i.e., two rounds).

In the next paragraphs we present how the read and write operation 
latency is affected by the scenarios we discussed in Section \ref{ssec:plscenarios}.
A general conclusion extracted from the experiments is that 
in most of the runs, algorithms \aprxsfw{} and \cwfr{} perform better than 
algorithm \simple{}. This suggests that the additional computation 
incurred in these two algorithms does not exceed the delay
associated with a second communication round.

As mentioned before, in general, PlanetLab machines often go offline 
unexpectedly preventing some scenarios to finish. Due to this fact, 
there are some cases in the plots where we do not obtain 
any data for a particular scenario. As those cases are not 
frequent we ignore them in our conclusions below.

\begin{figure*}[!ht]
{\bf 40 Writers:}\\	
	\begin{center}
	\vspace{-.7cm}
	\includegraphics[totalheight=2.3in]{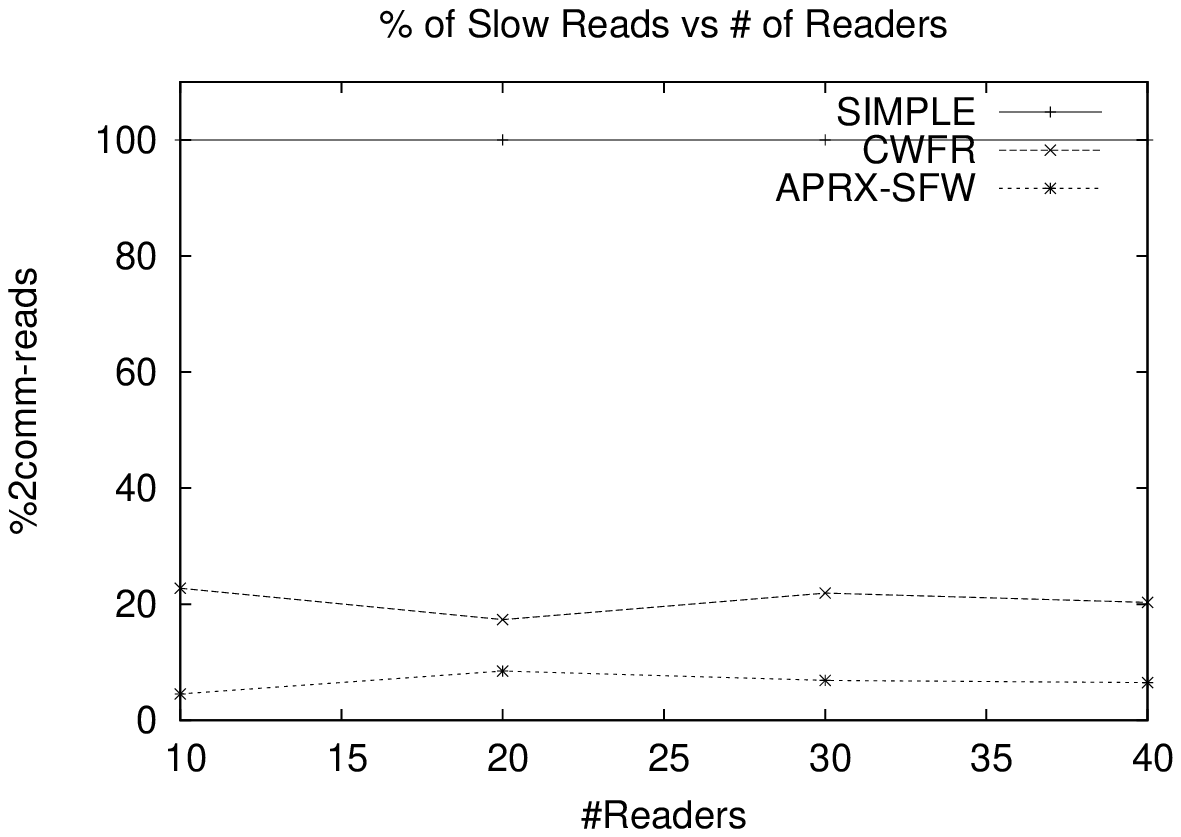}
	\hfill
	\includegraphics[totalheight=2.3in]{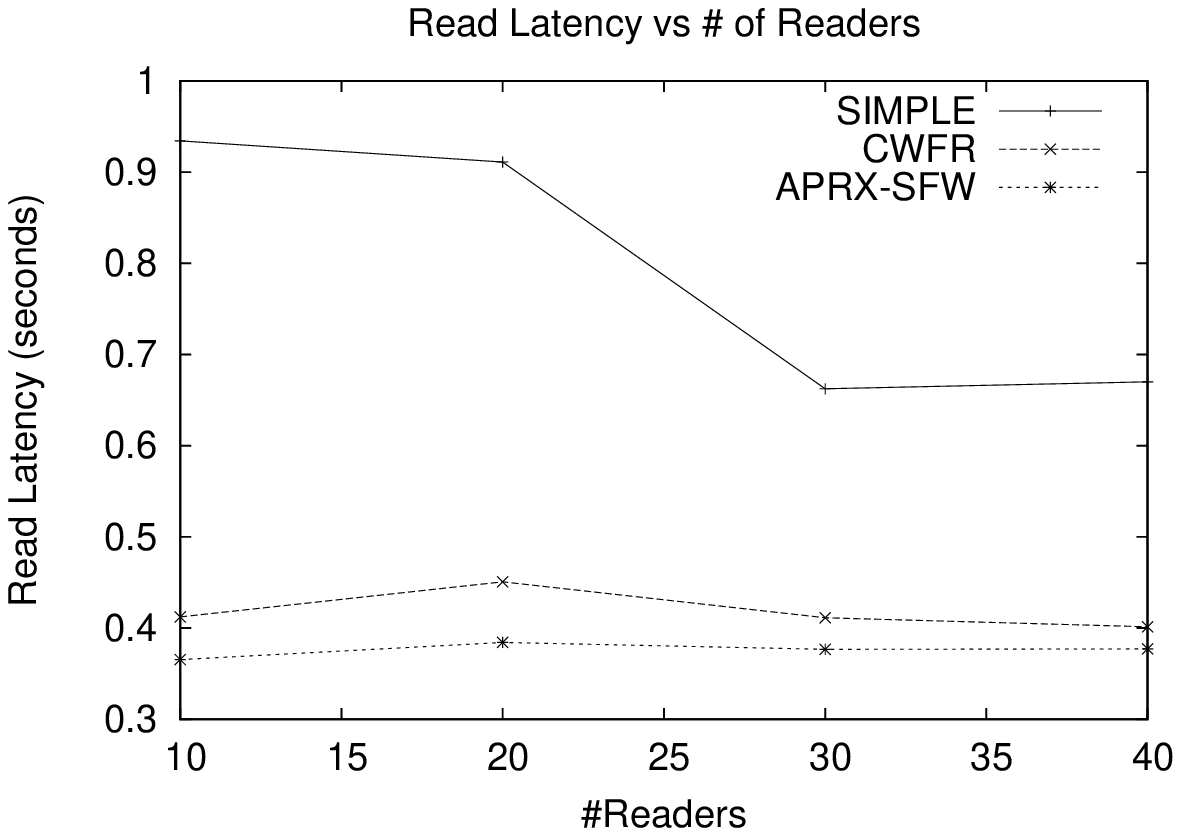}\\
	\end{center}
{\bf 40 Writers:}\\	
	\begin{center}
	\vspace{-.7cm}
	\includegraphics[totalheight=2.3in]{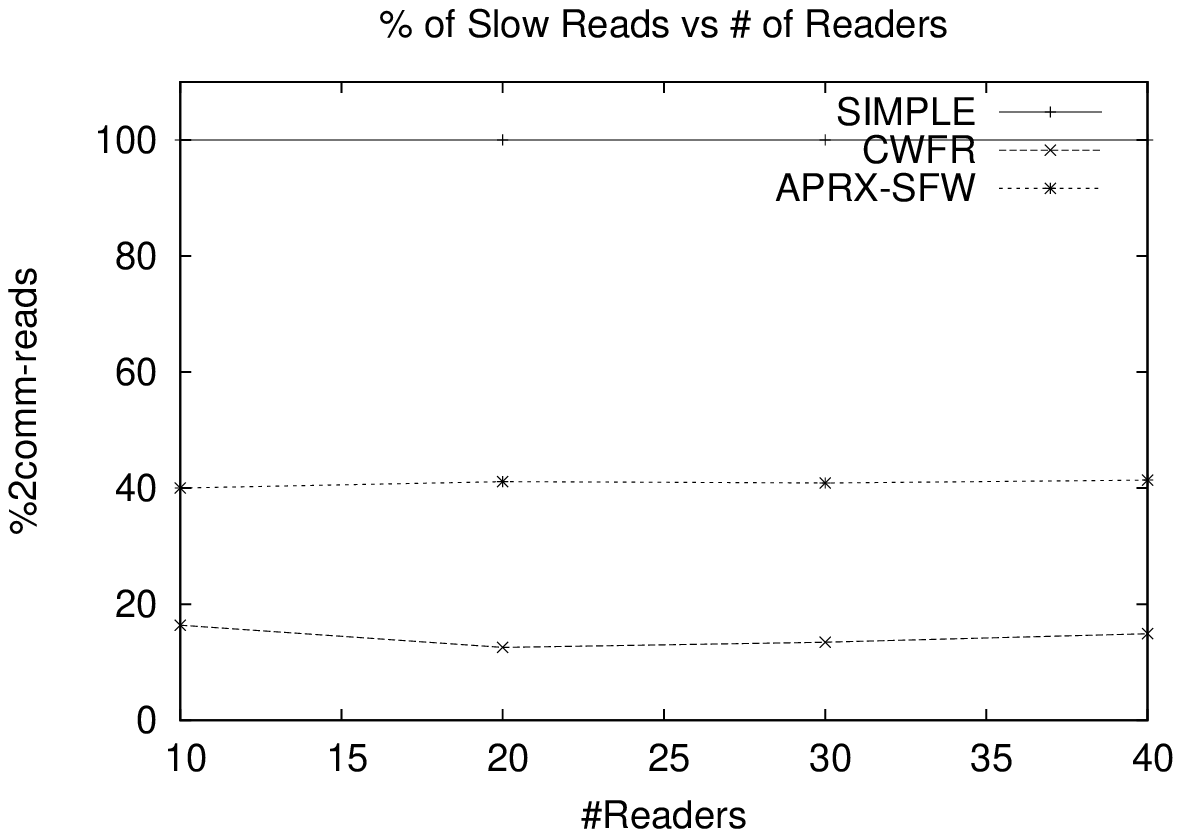}
	\hfill
	\includegraphics[totalheight=2.3in]{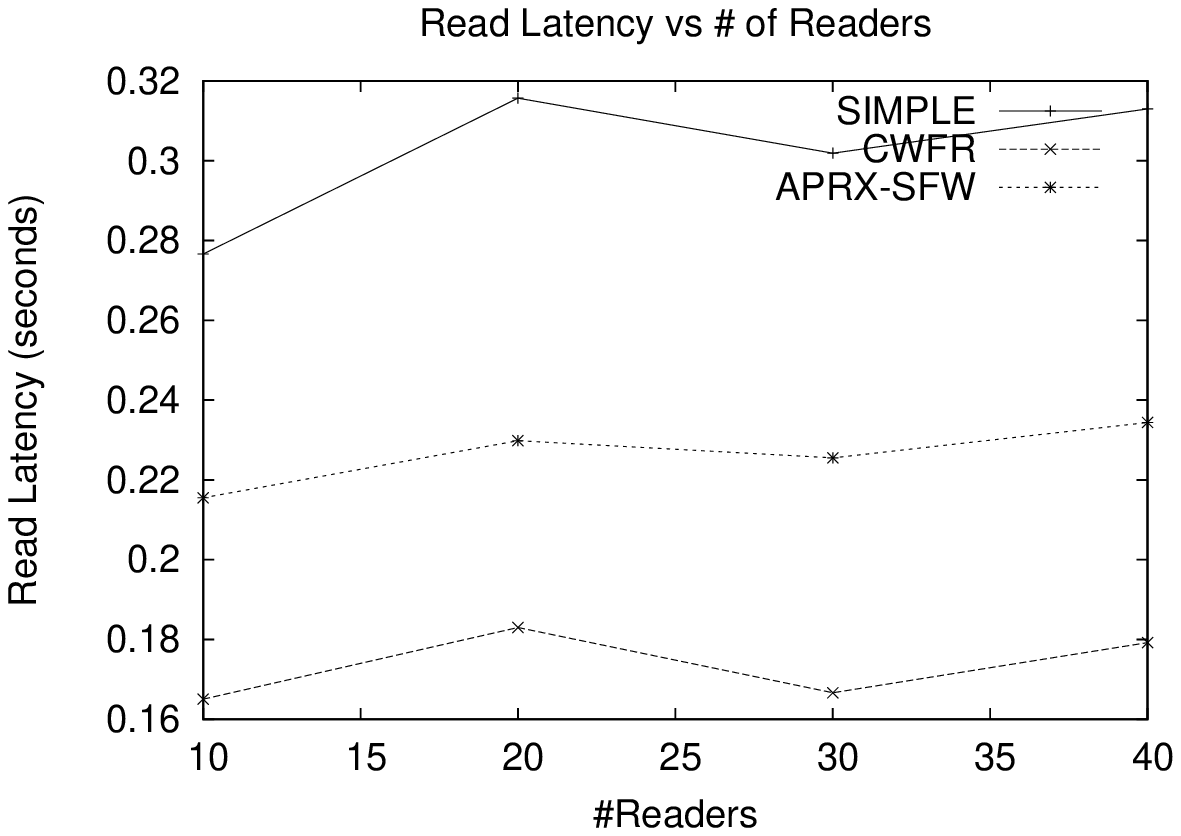}\\
	\end{center}
	\caption{ 
	19-wise quorum system ($|\srvSet=20$, $f=1$): {\bf Left Column:} Percentage of slow reads,  {\bf Right Column:} Latency of read operations }	
	\label{fig:plreadex}

\end{figure*}

\begin{figure*}[!ht]
{\bf 20 Readers:}\\	
	\begin{center}
	\vspace{-.7cm}
	\includegraphics[totalheight=2.3in]{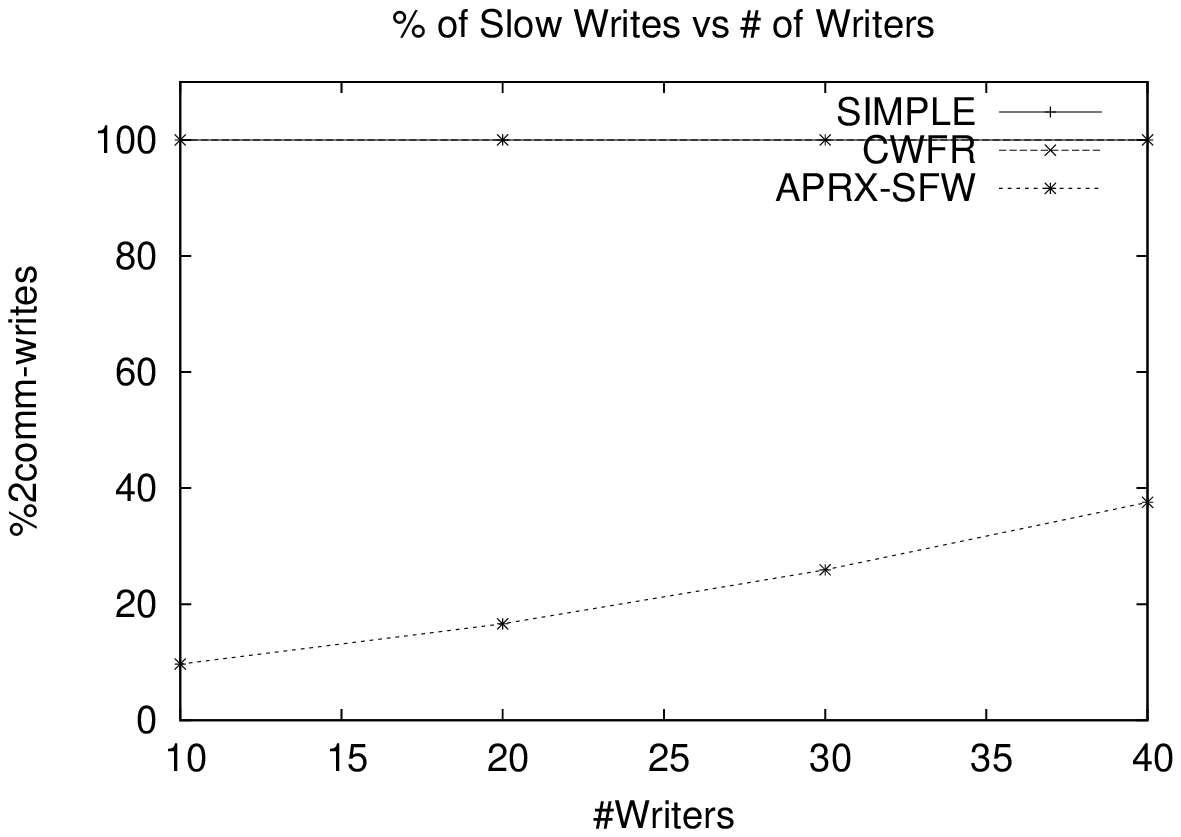}
	\hfill
	\includegraphics[totalheight=2.3in]{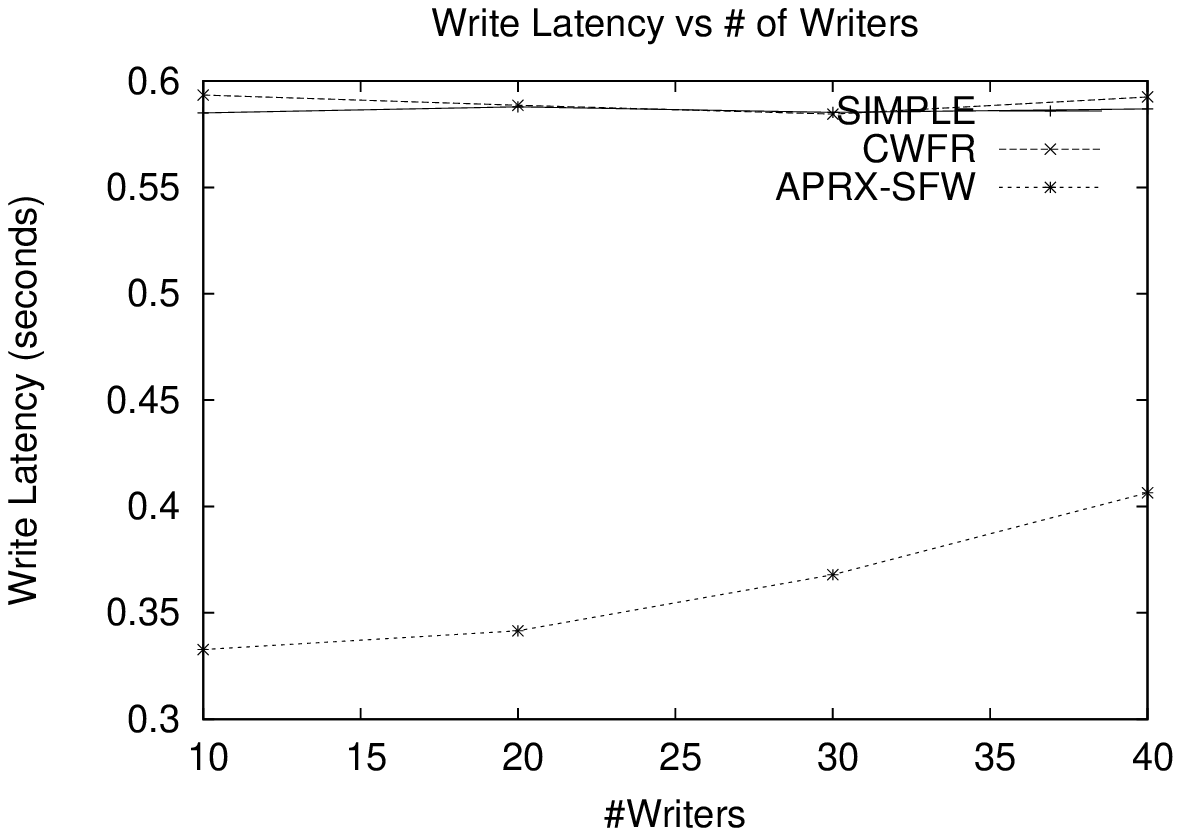}\\
	\end{center}
{\bf 20 Readers:}\\	
	\begin{center}
	\vspace{-.7cm}
	\includegraphics[totalheight=2.3in]{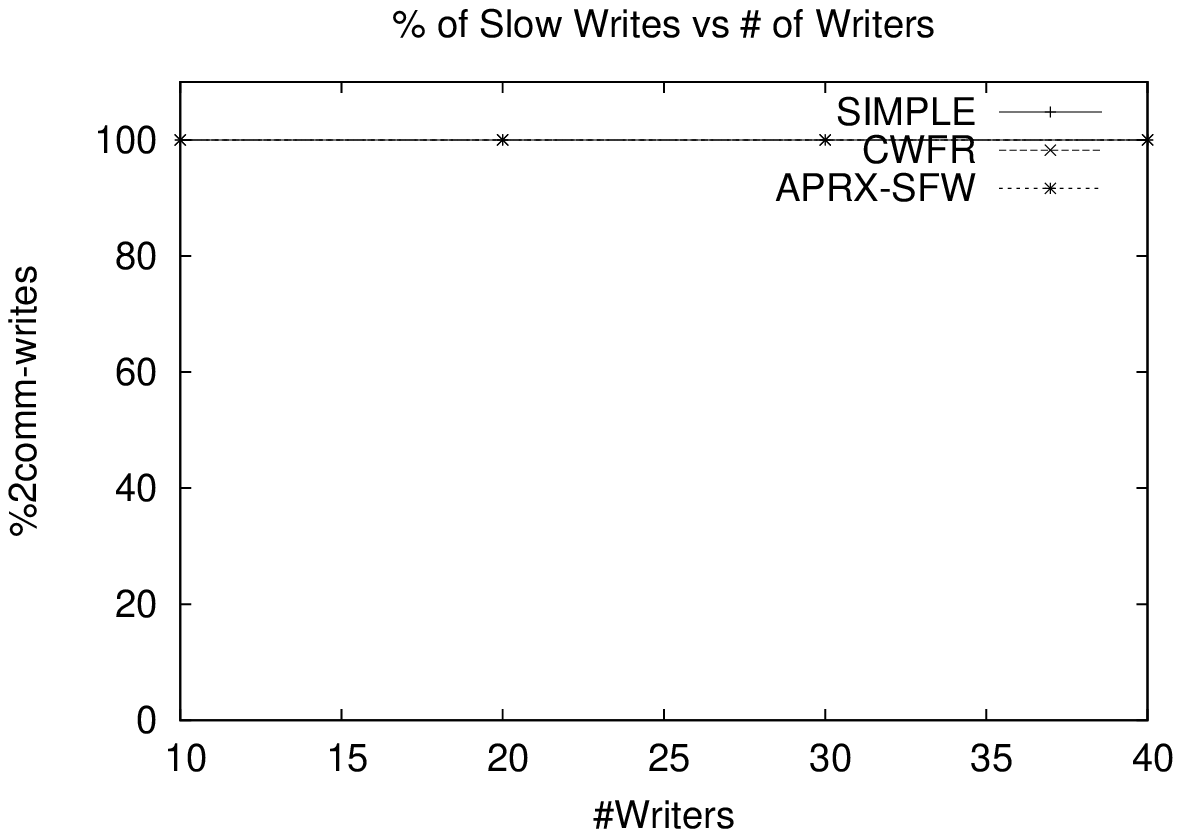}
	\hfill
	\includegraphics[totalheight=2.3in]{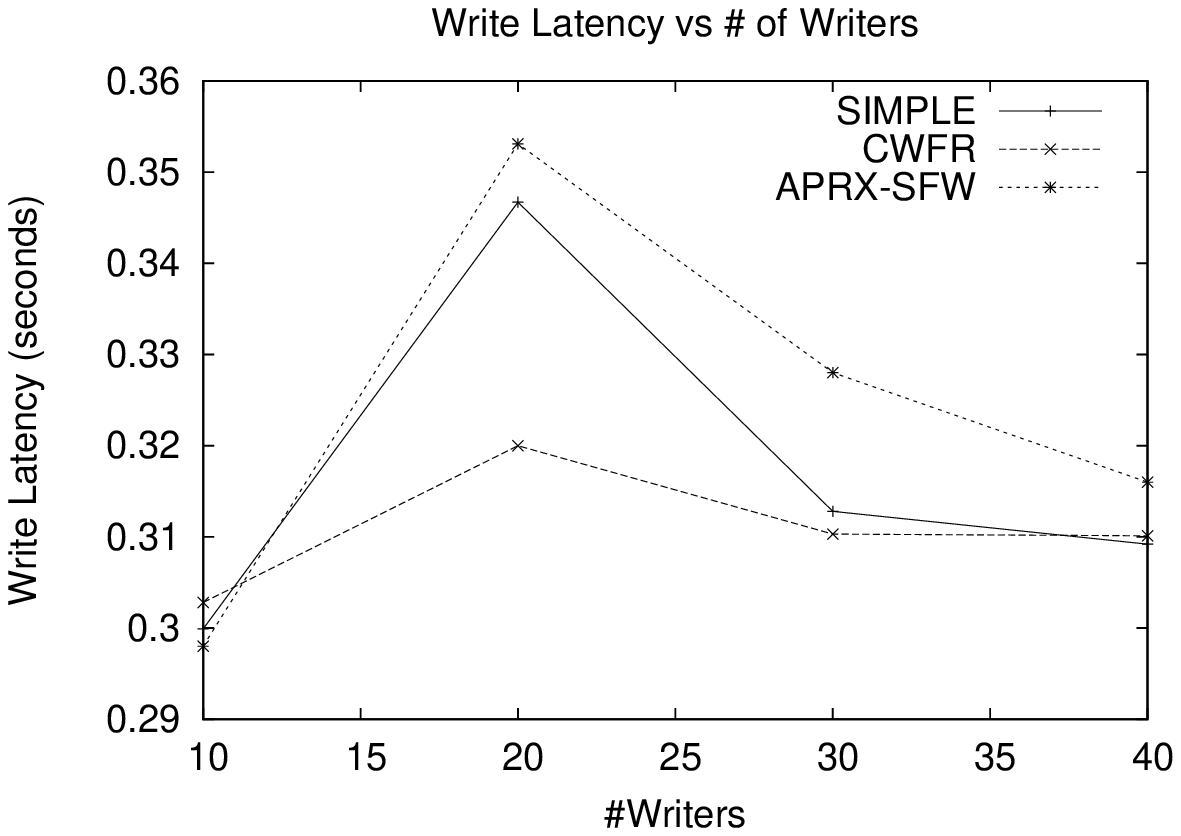}\\
	\end{center}
	\caption{9-wise and 4-wise quorum systems ($|\srvSet|=10$, $f=1,2$): {\bf Left Column:} Percentage of slow writes,  {\bf Right Column:} Latency of write operations}	
	\label{fig:plwriteex}
\end{figure*}

\paragraph{Variable Participation}
For this family of scenarios we tested the scalability of the algorithms 
when the number of readers, writers, and servers changes. 
The plots that appear in Figures \ref{fig:plreadex} and \ref{fig:plwriteex}
present the results 
for the read and write latency respectively.

Figure \ref{fig:plreadex} presents an example on how the performance of read
operations is affected as we change the number of readers. 
To obtain the figure we fix the number of 
servers and server failures, and we vary the number of readers and writers. 
So, for instance, Figure \ref{fig:plreadex} fixes the number of 
servers to $|\srvSet|=10$, and $f=1$. Each row of the figure fixes the 
number of writers and varies the number of readers. Therefore we obtain four 
rows corresponding to $|\wSet|\in[10,20,30, 40]$. 
Each row in a figure contains a pair of plots that presents the percentage 
 of slow reads (on the left) and the latency of each read (on the right) 
as we increase the number of readers. 

From the plots we observe that, unlike the results we obtained on the controlled environment 
of NS2, \aprxsfw{} over-performs \cwfr in most of the cases. 
This makes it evident that the adverse environment of a real-time system 
diminishes the high concurrency between read and write operations.
Notice that \aprxsfw{} and \cwfr{} require higher computation demands 
when multiple writers try concurrently to change the value of the register.
Computation is decreased when writes are consecutive. Thus, low concurrency
favors algorithms with high computation and low communication demands. 

The read performance of both \aprxsfw{} and \cwfr{} is affected slightly 
by the number of readers in the system. On the other hand we observe
that the number of writers seem to have a greater impact on the performance of algorithm 
\cwfr. As the number of writers grows,  
both the number of slow read operations and the latency of read operations for 
\cwfr{} increase. Unlike our findings in NS2 simulation, the number of writers does not have a great
impact on the performance of \aprxsfw{}. This is another evidence that write operations
do not overlap due to the adverse environment, and thus \aprxsfw{} can discover quickly 
a valid tag without the need of examining all the candidate tags. 
Both \cwfr{} and \aprxsfw{} require 
fewer than 20\% of reads to be slow in all scenarios. The only scenario where the 
percentage of slow reads for \aprxsfw{} rise above 50\% is when we deploy a 4-wise quorum 
system. This demonstrates the dependence of the predicates on the intersection degree 
of the underlying quorum system. We discuss this relation in the next paragraph.  
The large percentage of fast read operations, keeps the overall latency of read
operations for the two algorithms below the latency of read operations in \simple{}. That 
suggests that the computation burden does not exceed the latency added by a second communication 
round. 


As can be seen in Figure \ref{fig:plreadex}, the 20-wise 
intersection allows \aprxsfw{} to enable more single round read operations
than \cwfr{} even if there are 40 writers in the service.
The 4-wise intersection, of the plots on the second row of Figure 
\ref{fig:plreadex} causes the \aprxsfw{} to 
introduce a higher percentage of slow reads. In this case the 
predicates could not get validated and thus many reads needed to 
perform two communication rounds. \cwfr{} allows fewer slow reads than 
\aprxsfw{}. As expected, the read latency of \cwfr{} in this scenario 
is less than the read latency of \aprxsfw{}. 
Despite the higher percentage of slow reads however, the read latency of 
both algorithms remains below the read latency of \simple{}.
 
 
Similar to the read operations, the performance of write operations is not affected 
by the number of reader participants. It is affected however by both the number of
writers and servers in the system. The only algorithm that allows single round 
write operations is \aprxsfw{}. This is very distinct in our experiments 
and can also be seen in the first row of Figure \ref{fig:plwriteex}. 
When deploying a 9-wise quorum system \aprxsfw{} allows more than 60\% of 
the writes to be fast. We observe that the percentage of slow writes increases 
as the number of writers grows. This is expected as more write operations may collide and 
thus more tags are going to be propagated in the system. It is a pleasant surprise that 
the latency of write operations of \aprxsfw{} is below the write latency of the algorithms
that require two communication rounds. This is again an evidence that the computation
burden of \aprxsfw{} does not exceed the time of the second communication round. 
Note here that unlike the read operations, writes can be fast in \aprxsfw{} only when the write predicate holds. 
This characteristic can be depicted from the second row of Figure \ref{fig:plwriteex}
 where the 4-wise intersection does not allow 
for the write predicate to hold. Thus, every write operation in \aprxsfw{} performs
two communication rounds in this case. It is of great importance that the write latency of 
\aprxsfw{} is almost identical to the delay of the other two algorithms.
That shows once more that the computation burden of \aprxsfw{} does not affect the 
latency of writes by a high margin. The spikes on the latency of the write operations 
in the same figure appear due to the small range of the values and the small time inconsistency 
that may be caused by the variable delays of the network. 

\paragraph{Quorum Construction}
We consider majority quorums due to the property 
they offer on their intersection degree \cite{EGMNS09}:
if $|\srvSet|$ the number of servers and up to $f$ of them may crash
then if every quorum has size $|\srvSet|-f$ we can construct
a quorum system with intersection degree $n=\frac{|\srvSet|}{f}-1$.
Using that property we obtain the quorum systems presented on Table \ref{tab:quorums} 
by modifying the number of servers and the maximum number of server failures.

\begin{figure*}[!ht]
{\bf 20 Readers, 30 Writers, f=2:}\\	
	\begin{center}
	\vspace{-.7cm}
	\includegraphics[totalheight=2.3in]{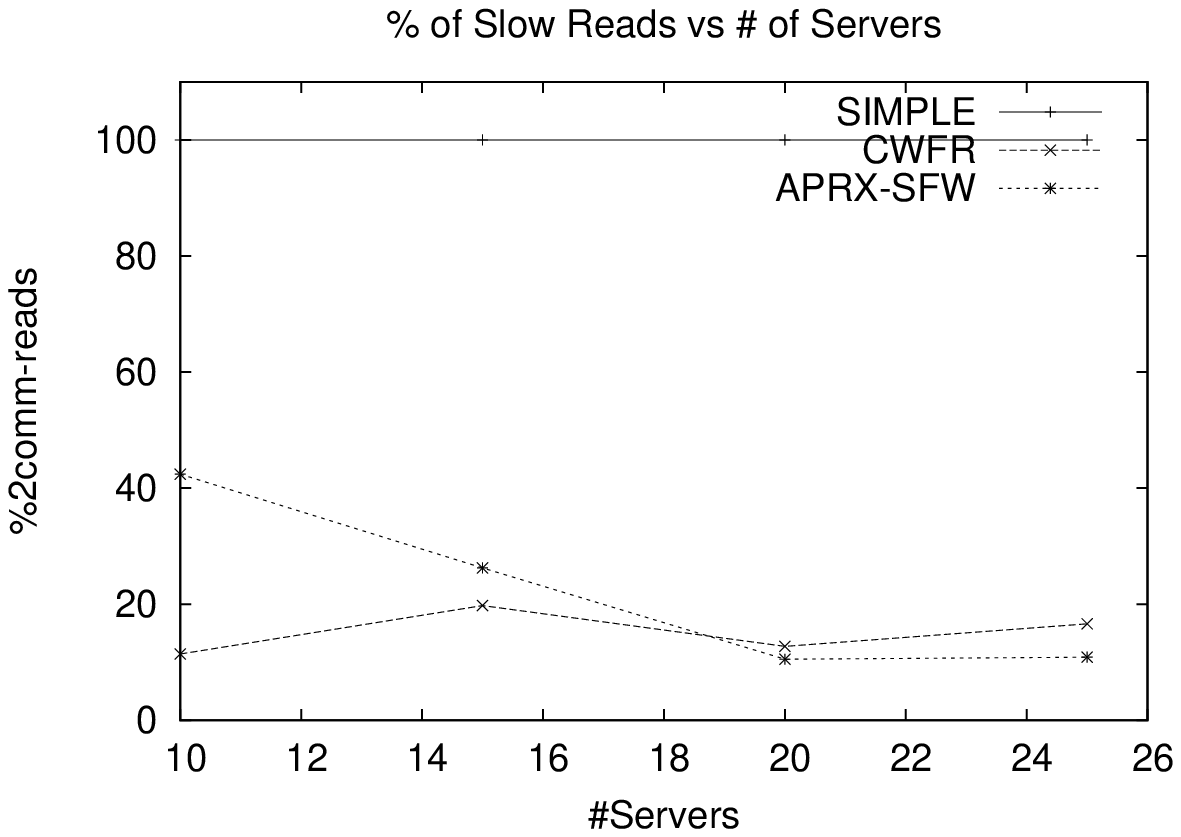}
	\hfill
	\includegraphics[totalheight=2.3in]{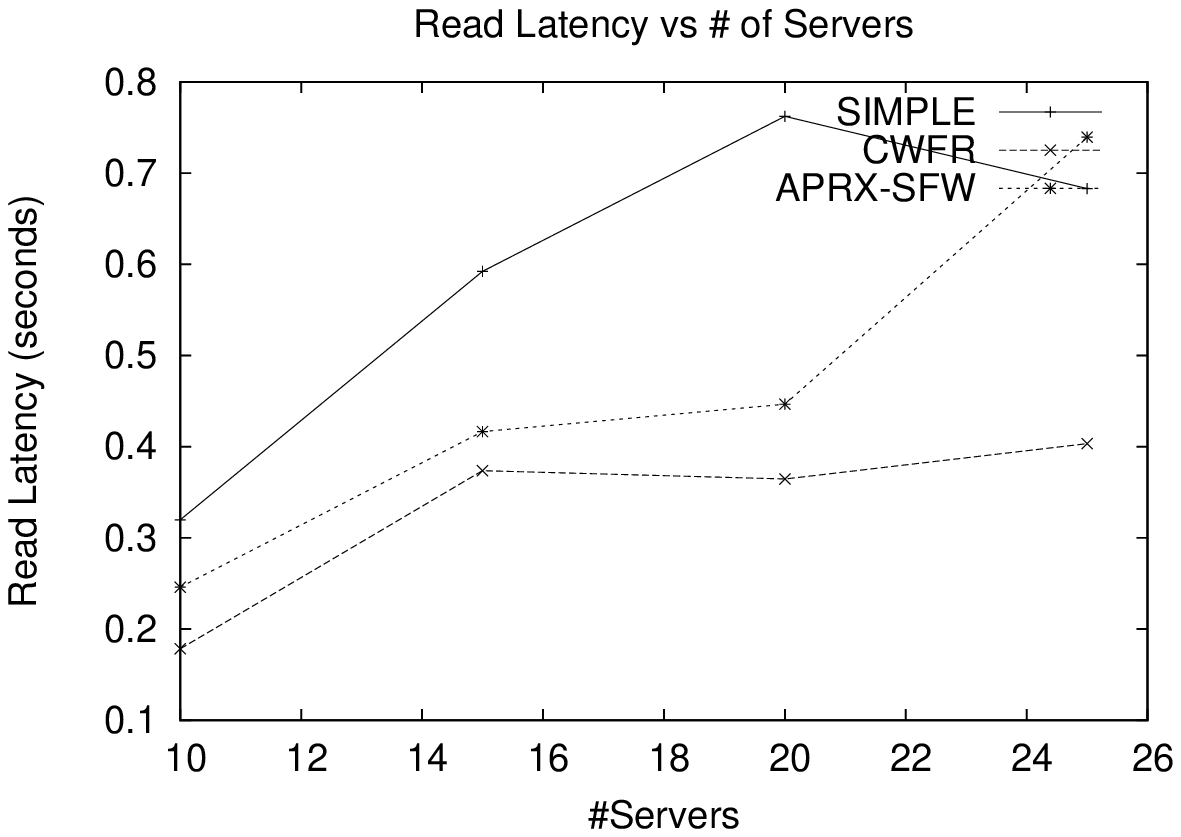}\\
	\includegraphics[totalheight=2.3in]{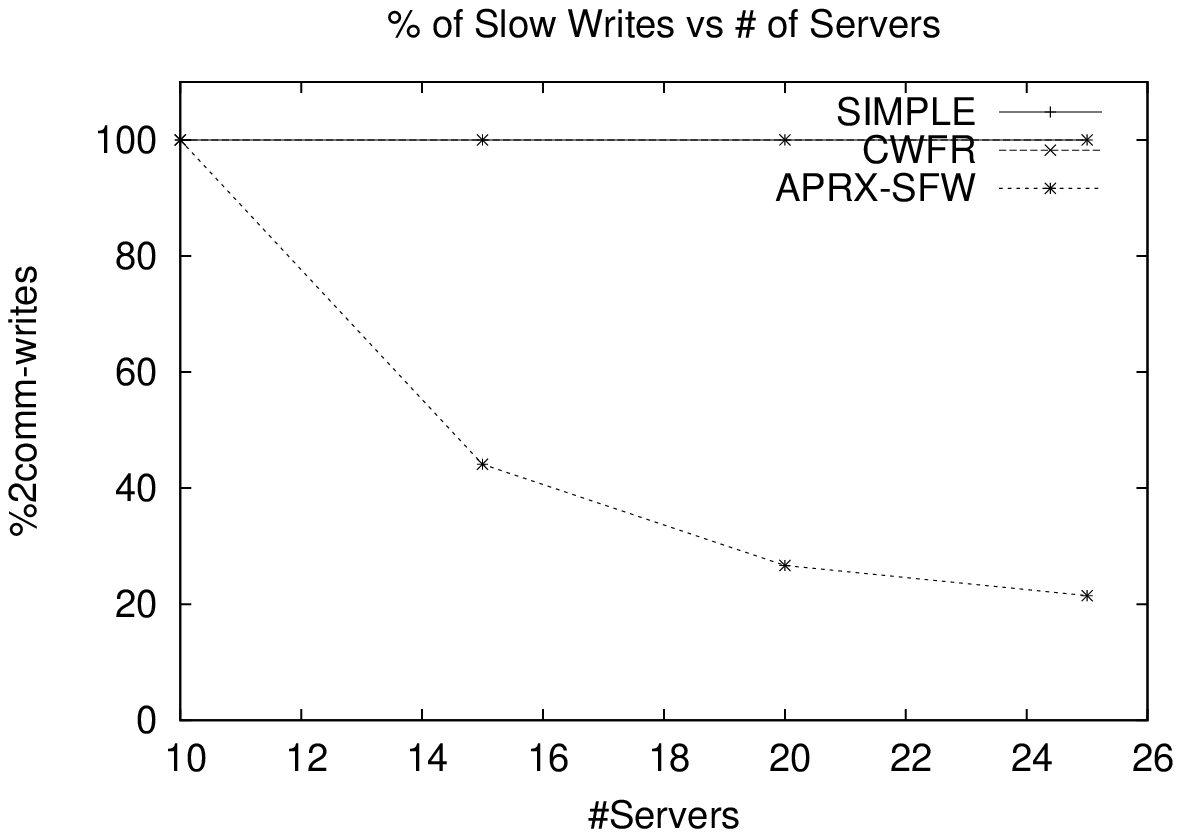}
	\hfill
	\includegraphics[totalheight=2.3in]{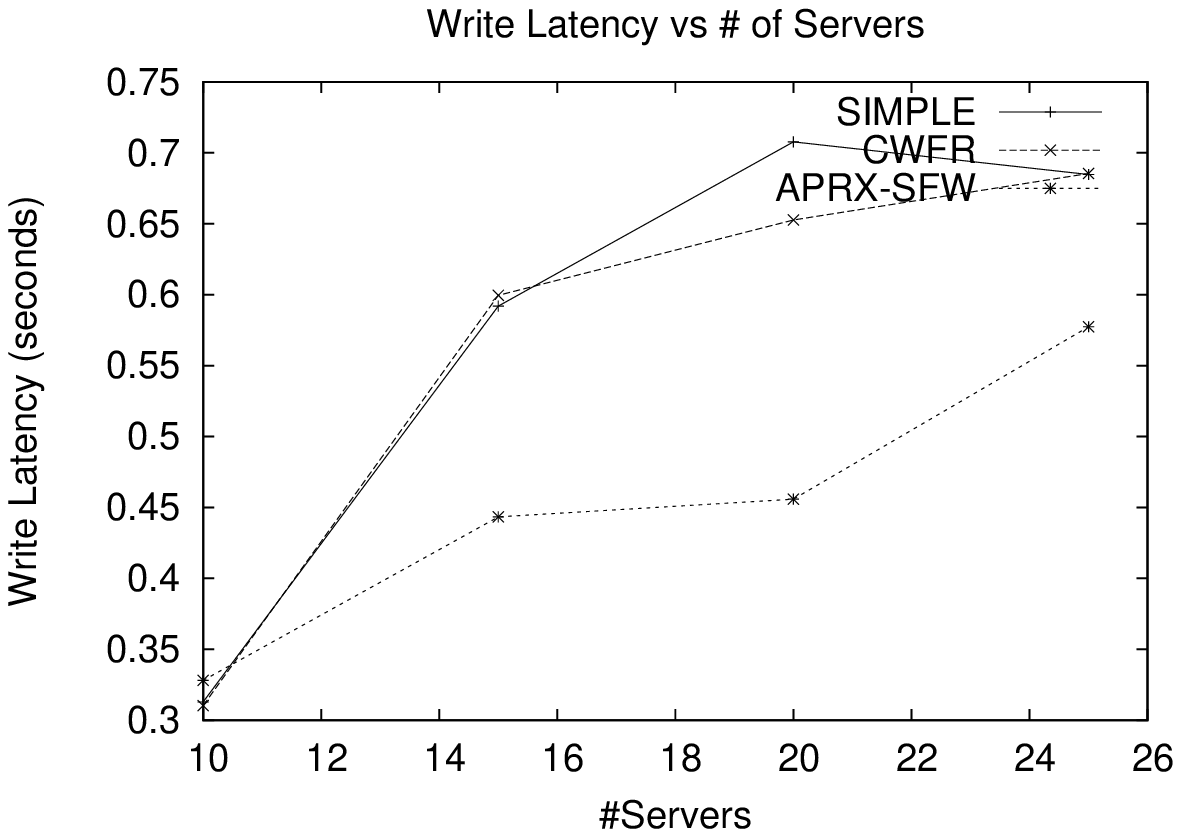}\\
	\end{center}
	\caption{{\bf Left Column:} Percentage of slow operations,  {\bf Right Column:} Latency of operations}	
	\label{fig:plquorumsex}
\end{figure*}

Figure \ref{fig:plquorumsex} illustrates an example of the performance of 
read and write operations (communication rounds and latency) with 
respect to the number of quorum members in the quorum system.
The two pairs that appear in the figure correspond to the quorum system that allows 
up to two server failures. The top pair
describes the performance of read operations while the bottom pair
the performance of write operations. Lastly, the left plot in each pair 
presents the percentage of slow read/write operations, and the right
plot the latency of each operation respectively.

From the plots it is clear that \aprxsfw{} is affected by the 
number of quorums and in extent by the intersection degree of 
the quorum system. 
In particular, we observe that the  cardinality of the quorum system, 
reduces the percentage of slow reads for 
\aprxsfw{}. On the other hand \cwfr{} does not seem to be affected. 
The distinction between \aprxsfw{} and \cwfr{} is depicted nicely 
from the upper row of Figure \ref{fig:plquorumsex}. The line of 
the slow read percentage of \aprxsfw{} crosses below the line of \cwfr{} 
when the intersection degree grows to $\frac{20}{2}-1=9$, and remains
lower for larger values of the intersection degree. 
Operation latency on the other hand is not proportional to 
the reduction on the amount of slow operations. Both algorithms 
\aprxsfw{} and \cwfr{} experience an incrementing trend on the latency of 
the read operations as the number of servers and the quorum members
increases. It is worth noting that the latency of read operations in \simple{} 
also follows an increasing trend even though every read operation requires
two communication rounds to complete. This is evidence that the increase on the 
latency is partially caused by the communication between the readers and the servers:
as the servers increase in number the readers need to send and wait for more messages.
The latency of read operations in \cwfr{} is not affected greatly as the 
number of servers changes. 
As a result, \cwfr{} appears to maintain a read latency close to 0.4 sec in 
every scenario. With this read latency \cwfr{} over-performs algorithm \simple{}
in every scenario as the latter maintains a read latency between 0.6 and 1 sec.

Similar observations can be made for the write operations. 
Observe that although both \cwfr{} and \simple{} require 
two communication rounds per write operation, the write 
latency in these algorithms is affected negatively by the 
increase of the number of quorums in the system. As we said before
this is evidence of the higher communication demands when we increase
the number of servers. We also note that the latency of the write operations
in \simple{} is almost identical to the latency of the read operations of
the same algorithm. This proves the fact that the computation demands in either 
operation is also identical. As for \aprxsfw{} we observe that the increase 
on the number of servers reduces the amount of slow write operations 
as also observed during the NS2 simulation. Also we observe 
that the average write latency of \aprxsfw{} 
increases as the number of quorums increases in the system. Interestingly 
however, unlike the findings in \cite{D5-TR}, the latency of writes does remain 
in most cases below the competition. Comparing with 
the latency of read operations it appears that although \aprxsfw{} may allow 
more fast read operations than writes under the same conditions, 
the average latency of each read is almost the same as the latency of write operations.
An example of this behavior can be seen in Figure \ref{fig:plquorumsex}.
In this example slow reads can drop as low as 10\%  and the average read latency
climbs to an average of 0.5 seconds (the 0.7 sec point appears to be an outlier). 
On the other no less than 20\% slow writes are allowed
and the average write latency climbs just above 0.5 seconds. The simple explanation for 
this behavior lies on the evaluation of the read and write predicates: each reader needs
to examine the latest tags assigned to every writer in the system whereas each writer only 
examines the tags assigns to its own write operation.


\section{Conclusions}
\label{sec:conc}
This work investigates the efficiency and practicality of four MWMR atomic register algorithms
designed for the asynchronous, failure prone, message passing environment. 
The experiments involved the implementation and evaluation of the algorithms 
on the NS2 single-processor simulator and on the planetary-scale PlanetLab platform.
We first provided a proof-of-concept comparison of \sfw{} with \aprxsfw{}
on the NS2 simulator. Then we compared the performance of algorithms \cwfr{}, \aprxsfw{} and
\simple{} on both NS2 and on PlanetLab. NS2 allowed us to 
test the algorithms under specific environmental conditions. 
In contrary, in the adverse real-time environment offered by PlanetLab, we were able to 
observe the behavior of the algorithms in real-time network conditions. 
Our tests in both environments involved  the scalability of the algorithms using variable number of participants
and the impact that the deployed quorum system may have on each algorithm's performance.
In the NS2 simulator we were also able to test how the network delay may affect the 
performance of the algorithms. 

The results for \cwfr{}, \aprxsfw{} and \simple{}, suggested that algorithms \cwfr{} and
\aprxsfw{} over-perform \simple{} in both NS2 and PlanetLab.  We also observed that 
the number of writers, servers and quorums in the system can have a negative impact 
on both the number of slow operations, and the operation latency for all three algorithms. 
This behavior agrees with our theoretical bounds presented in Table \ref{tab:rev}.
Algorithm \cwfr{} does not seem to be affected a lot by the number of servers and quorums in the
underlying quorum system. The percentage of slow reads remains in the same levels 
while the latency increases due mainly to the increasing communication. 
 
A factor that favors \aprxsfw{} is the intersection degree of the 
underlying quorum system. This was more evident in the PlanetLab implementation 
and less in the NS2 simulation. 
Unlike the NS2 simulation, we observed that in a real-time environment \aprxsfw{} 
over-performs \cwfr{} 
in most scenarios were the intersection degree is higher than 6.  That is, both 
the percentage of slow reads along with the average read latency is lower than 
the read performance of \cwfr{}.  
However, it appears that \aprxsfw{} performs well in cases with small intersection degree. 
Less than 50\% of reads
needed to be slow and the latency of both the write and read operations was almost identical 
to the latency of the competition when assuming a 4-wise quorum system. 

Finally, the network delay has a negative 
impact on the operation latency of every algorithm we tested. 
We observed that network delays promote in some cases 
the use of algorithms with high computation demands that minimize the communication 
rounds, like algorithm \aprxsfw{}. This adheres with the observed performance
of algorithm \aprxsfw{} in PlanetLab, which has relatively large network delay (when
compared for example with a local area network). 

In general, our results suggest that the computation burden placed on each participant 
in algorithms \aprxsfw{} and \cwfr{} is less than the cost of a second communication round. 
The computation cost is not negligible however since we observe cases where the 
percentage of slow reads decreases when the average read latency increases. 
This reveals that although we need less communication, the computation 
costs keep operation latency incrementing. 

Overall it is safe to claim that algorithms \aprxsfw{} and \cwfr{} should be preferable by 
applications in a real-time setting. The choice of the algorithm (\cwfr{} or \aprxsfw{}) 
depends significantly on the environment in which the application is to be deployed. 
If quorum systems with large intersection degrees are possible/available then \aprxsfw{} is a clear winner. 
If a general quorum construction is to be used then \cwfr{} should be considered.

An interesting future work would be to investigate how the examined algorithms scale 
with high throughput when participants may pipeline multiple requests. Also it is still 
an open question how the quorum construction may affect the performance of these 
algorithms. Although we provide preliminary results with different families of quorum 
systems we do not investigate how the server location or the network delays may lead
participants to communicate with specific quorums. Such information may allow for 
quorum constructions that further improve operation latency.  

\paragraph{Acknowledgments.} We thank Alexander Russell and Alexander A. Shvartsman
for helpful discussions.

\bibliographystyle{acm}
\bibliography{biblio}

\pagebreak

\end{document}